\documentclass[sigconf]{acmart}

\usepackage{paralist}
\usepackage{subfig}
\usepackage{caption}
\usepackage[export]{adjustbox}
\usepackage{multirow}
\usepackage{bm}
\AtBeginDocument{%
  }

\copyrightyear{2022} 
\acmYear{2022} 
\setcopyright{acmcopyright}\acmConference[CIKM '22]{Proceedings of the 31st ACM International Conference on Information and Knowledge Management}{October 17--21, 2022}{Atlanta, GA, USA}
\acmBooktitle{Proceedings of the 31st ACM International Conference on Information and Knowledge Management (CIKM '22), October 17--21, 2022, Atlanta, GA, USA}
\acmPrice{15.00}
\acmDOI{10.1145/3511808.3557546}
\acmISBN{978-1-4503-9236-5/22/10}

\begin{document}

\title{An Empirical Study on the Membership Inference Attack against Tabular Data Synthesis Models}


\author{Jihyeon Hyeong}
\email{jihyeon.hyeong@northwestern.edu}
\affiliation{%
  \institution{Northwestern University}
  \country{United States}}

\author{Jayoung Kim}
\email{jayoung.kim@yonsei.ac.kr}
\affiliation{%
  \institution{Yonsei University}
  \country{South Korea}}

\author{Noseong Park}
\email{noseong@yonsei.ac.kr}
\affiliation{%
  \institution{Yonsei University}
  \country{South Korea}}

\author{Sushil Jajodia}
\email{jajodia@gmu.edu}
\affiliation{%
  \institution{George Mason University}
  \country{United States}}

\renewcommand{\shortauthors}{Hyeong, et al.}

\begin{abstract}
Tabular data typically contains private and important information; thus, precautions must be taken before they are shared with others. Although several methods (e.g., differential privacy and k-anonymity) have been proposed to prevent information leakage, in recent years, tabular data synthesis models have become popular because they can well trade-off between data utility and privacy. However, recent research has shown that generative models for image data are susceptible to the membership inference attack, which can determine whether a given record was used to train a victim synthesis model. In this paper, we investigate the membership inference attack in the context of tabular data synthesis. We conduct experiments on 4 state-of-the-art tabular data synthesis models under two attack scenarios (i.e., one black-box and one white-box attack), and find that the membership inference attack can seriously jeopardize these models. We next conduct experiments to evaluate how well two popular differentially-private deep learning training algorithms, DP-SGD and DP-GAN, can protect the models against the attack. Our key finding is that both algorithms can largely alleviate this threat by sacrificing the generation quality.
\end{abstract}

\ccsdesc[500]{Security and privacy~Management and querying of encrypted data}
\ccsdesc[500]{Security and privacy~Privacy protections}
\ccsdesc[500]{Security and privacy~Usability in security and privacy}
\keywords{tabular data synthesis; membership inference attack; differential privacy}

\maketitle
\section{Introduction}

Tabular data is one of the most popular data types in reality. It must be kept confidential when it includes valuable and sensitive information. To achieve this, many different approaches have been introduced to protect the privacy of tabular data~\cite{Jordon2019PATEGANGS, anonymization}. In recent years, tabular data synthesis using deep models has been actively investigated since it synthesizes fake records after learning from real records; synthesized tables show behavior similar to the original tables. Generative models guarantee enhanced privacy protection as they output fake records that do not directly correspond to any specific real record, unlike previous methods (e.g., k-anonymity~\cite{kanonymity} or differential privacy ~\cite{chen2018differentially}). However, recent research has revealed that generative models are susceptible to the membership inference attack (MIA), where attackers can infer if a given record was used to train a victim model~\cite{Dwork}. This can put its original training data at substantial risk of being leaked to the attackers.

Unfortunately, most of the research on MIA against generative models has focused on image data~\cite{hayes2018logan, hilprecht2019monte, Chen_2020}. In this paper, we empirically study MIA in the context of tabular data synthesis. Using the state-of-the-art MIA methods, in other words, we attack the state-of-the-art tabular data synthesis models, which have not been done previously. To our knowledge, we are the first to report the results. More specifically, we conduct experiments in which the tabular synthesis models trained on 4 tabular datasets are attacked under the following two scenarios: i) the black-box attack, where attackers have access only to fake records generated by the victim model, and ii) the white-box attack, where attackers know the internals of the victim model.

We evaluate and compare attack success scores under the two attack scenarios. Moreover, we analyze under what condition tabular data synthesis models become vulnerable to the attack. As a countermeasure to MIA, we conduct additional experiments where the victim models are trained with the two differentially private (DP) deep learning training algorithms -- DP-SGD~\cite{Abadi_2016} and DP-GAN~\cite{xu2018dpgan} -- and compare the results with the models trained without any DP algorithm. These analyses have not been studied yet under the context of tabular data synthesis. Our key findings are as follows:
\begin{compactenum}
    \item As the attacker collects more fake records, the attack is more successful, but there is a limit over which the success rate does not increase. 
    \item The earth mover's distance (EMD) between real and fake tabular data highly correlates to the attack success rate. It is an effective surrogate to predict whether fake tabular data is vulnerable to MIA or not (cf. Table~\ref{tab:correlation}).
    \item Tabular data synthesis models trained with DP-GAN show better generation quality than those with DP-SGD, which are even similar to the models without DP in some cases.
    \item DP-GAN sometimes preserves the models' generation quality yet is vulnerable to a white-box attack; on the other hand, DP-SGP is more robust to the white-box attack but sacrifices the models' generation quality.
    \item In general, we recommend \texttt{OCT-GAN} among the tested models since it is balanced between synthesis quality and privacy.
\end{compactenum}

We note that our main contribution in this paper is that we attack state-of-the-art tabular data synthesis methods with state-of-the-art MIA methods and derive several findings as listed above. We hope that our findings can help other followers who want to study the MIA in the tabular data synthesis context.

\section{Related Work \& Preliminaries}
We review various state-of-the-art tabular data synthesis models, MIA attack methods, and the differential privacy.

\subsection{Tabular Data Synthesis}
\emph{Tabular data synthesis models} generate realistic records by modelling a joint probability distribution of columns in a table. As standard generative models do, a \emph{generator} network takes as its input a noisy vector $\bm{z}$ and outputs a fake record. However, as tabular data is characterized by its high dimensionality with a mix of numeric and categorical data, tabular data synthesis models are more designed for handling various distributions and computational problems. For instance, \texttt{TVAE}~\cite{ishfaq2018tvae} is a variational autoencoder that effectively handles mixed types of features in tabular data. There are also many variants of GANs for tabular data synthesis.  \texttt{TableGAN}~\cite{DBLP:journals/corr/abs-1806-03384} generates tabular data using convolutional neural networks. \texttt{CTGAN}~\cite{NIPS2019_8953}, well-known for its tabular data synthesis quality, takes a conditional generator and incorporates its mode-specific normalization process. \texttt{OCT-GAN}~\cite{kim2021octgan} is a model based on neural ordinary differential equations both in generator and discriminator. 

\subsection{Membership Inference Attack}
The \emph{membership inference attack} (MIA) against generative models aims to determine whether a given record was used to train the victim model~\cite{Dwork}. It can threaten privacy as not only does it disclose raw personal data used in training but also allows potential information leakage of the model. Shokri et al.~\cite{shokri2017membership} presented the first MIA against classification models in the black-box setting and quantified the membership information leakage. This led to increasing research on MIA in machine learning models, especially on generative models. Hayes et al.~\cite{hayes2018logan} proposed the first MIA against generative models for images both in the black and white-box settings. Hilprechet et al.~\cite{hilprecht2019monte} proposed two MIAs, one is for GANs in the black-box setting, and the other is for VAEs in the white-box one. Chen et al.~\cite{Chen_2020} proposed a generic attack model which comprehends all scenarios from the black-box to the white-box setting, which we adopt for our experiments as follows: 
\begin{enumerate}
    \item As generative models are trained to estimate the training data distribution, attackers infer the membership $\bm{m}_i$ by using the probability of a given data $\bm{x}$ being generated by the victim model $\mathcal{G}_v$ as follows:
    \begin{align}
        P(\bm{m}_i=1|\bm{x}_i,\theta_v)\propto P_{\mathcal{G}_v}(\bm{x}|\theta_v).
    \end{align}
    However, as calculating the exact probability is intractable, they approximate it using the Parzen window density estimation~\cite{DudaHartStork01}:
    \begin{align} \label{eq:Parzen}
        P_{\mathcal{G}_v}(\bm{x}|\theta_v) \approx \frac{1}{k}\sum_{i=1}^k \exp(-\|\bm{x}- \mathcal{G}_v(\bm{z}_i)\|_2).
    \end{align}
    \item In the black-box scenario, attackers blindly collect $k$ fake records generated by the victim model and get a reconstructed copy of a data $\bm{x}$ as $\mathcal{R}(\bm{x}|\mathcal{G}_v)$ to approximate the probability in Eq.~\eqref{eq:Parzen} as follows:
    \begin{align}
        \mathcal{R}(\bm{x}|\mathcal{G}_v) = \underset{\hat{\bm{x}}\in\{\mathcal{G}_v(\cdot)_i\}_{i=1}^k}{\arg\min} \|\bm{x} - \hat{\bm{x}}\|_2, 
    \end{align}
    where $\{\mathcal{G}_v(\cdot)_i\}_{i=1}^k$ is a set of $k$ fake records collected by attackers.
    \item In the white-box scenario, attackers get $\mathcal{R}(\bm{x}|\mathcal{G}_v)$ to compute the probability in Eq.~\eqref{eq:Parzen} through optimizing $\bm{z}$ as follows:
    \begin{align}
        \mathcal{R}(\bm{x}|\mathcal{G}_v) = \mathcal{G}_v(\bm{z}^*),
    \end{align}
    where $\bm{z}^* = \underset{\bm{z}}{\arg\min} \text{ }\|\bm{x} - \mathcal{G}_v(\bm{z})\|_2.$
\end{enumerate}
However, all previous research has focused on generative models for image data, not for tabular data, albeit some used tabular data in their experiments by heuristically modifying the models for tabular data. However, as tabular data requires pre-processing of its complex data distribution, such heuristics can mislead the analysis. To this end, this paper aims to target tabular data synthesis models for an accurate analysis on MIA on tabular data.

\subsection{Differential Privacy}
\emph{Differential privacy} (DP) is a stochastic defense mechanism for training machine learning models to preserve the membership privacy of individual samples against MIA~\cite{dwork2014algorithmic}. As a model trained with DP does not remember any specific record, DP has known to protect well machine learning models against MIA~\cite{Chen_2020, beaulieu2019privacy}. DP-SGD~\cite{Abadi_2016}, which perturbs the gradients of the model during training by clipping and adding noise to them, is the most widely-accepted DP algorithm. Adjusting DP-SGD being specialized for GANs, DP-GAN~\cite{xu2018dpgan} is another DP algorithm that achieves DP by replacing the gradient clipping with the weight clipping. 

\section{Empirical Study Design}
In this section, we describe our empirical study design. We conduct our experiments with 4 tabular data synthesis models and 4 real-world tabular datasets. 

\begin{figure*}
    \centering
    \captionsetup[subfloat]{justification=centering}
    
    \subfloat[\texttt{Adult}]{\includegraphics[width=0.48\columnwidth,trim={0.7cm 0 0 0},clip]{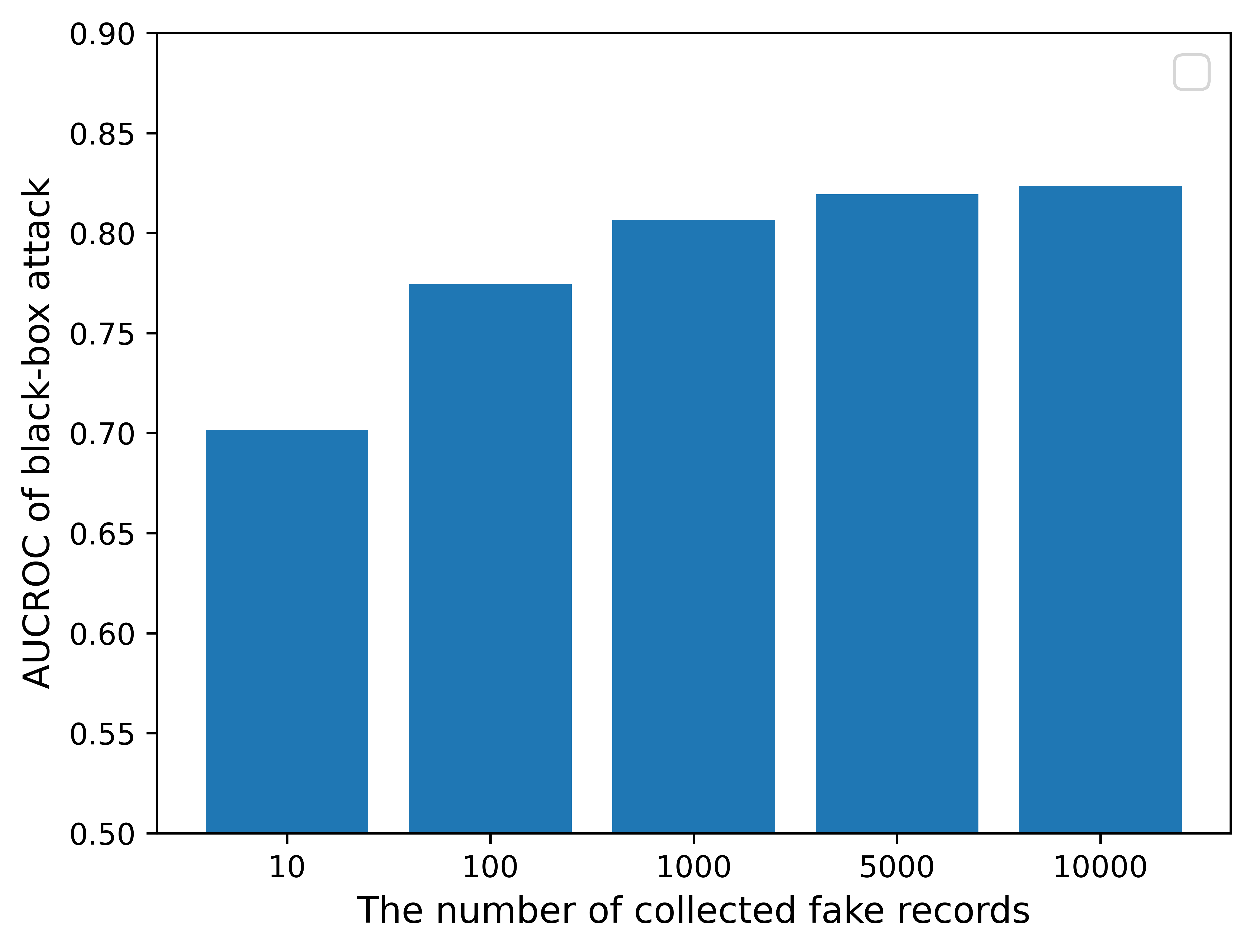}\label{fig4-a}} \hfill
    \subfloat[\texttt{Alphabank}]{\includegraphics[width=0.48\columnwidth,trim={0.7cm 0 0 0.24cm},clip]{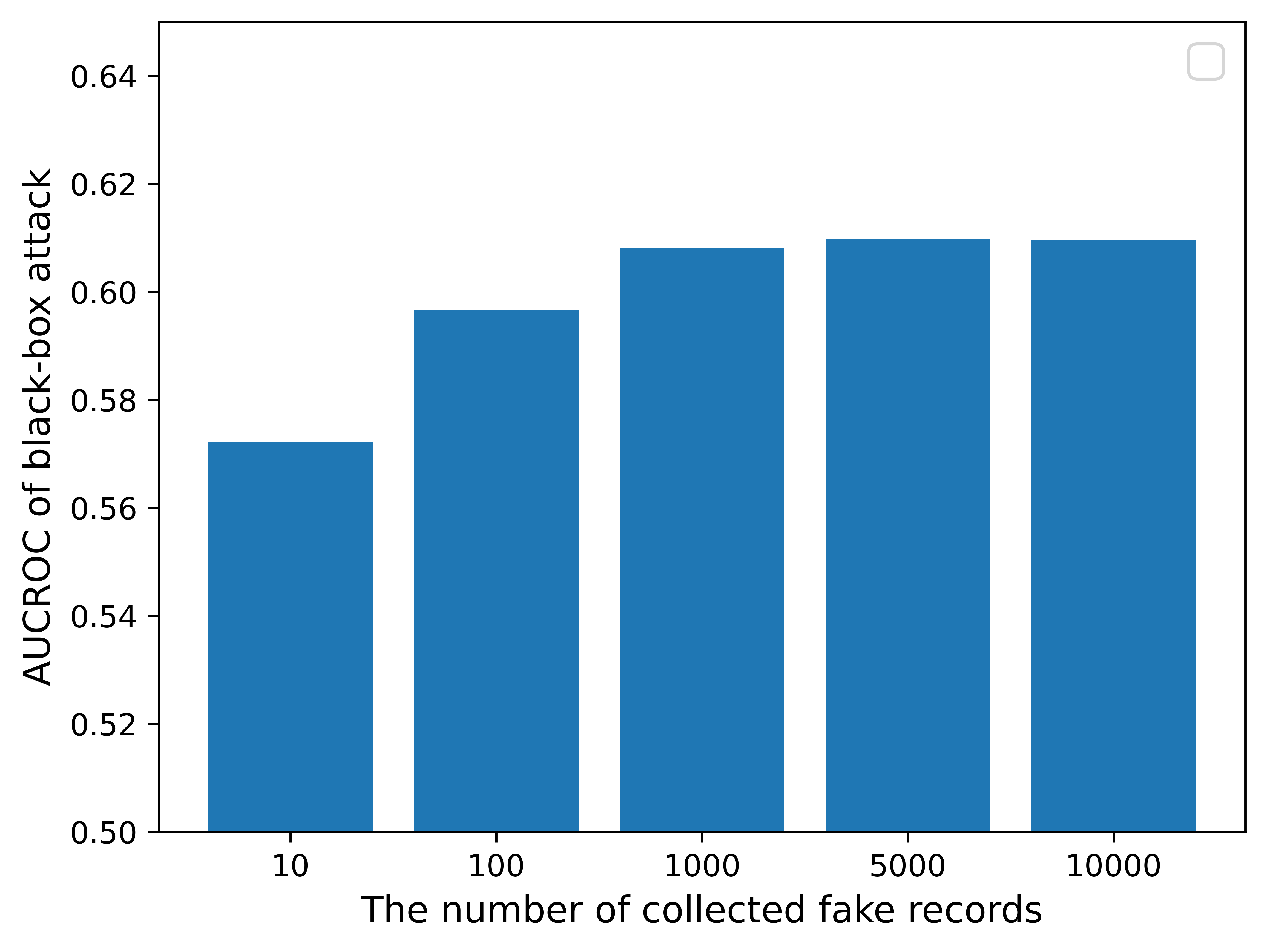}\label{fig1-b}} \hfill
    \subfloat[\texttt{Surgical}]{\includegraphics[width=0.48\columnwidth,trim={0.7cm 0 0 0},clip]{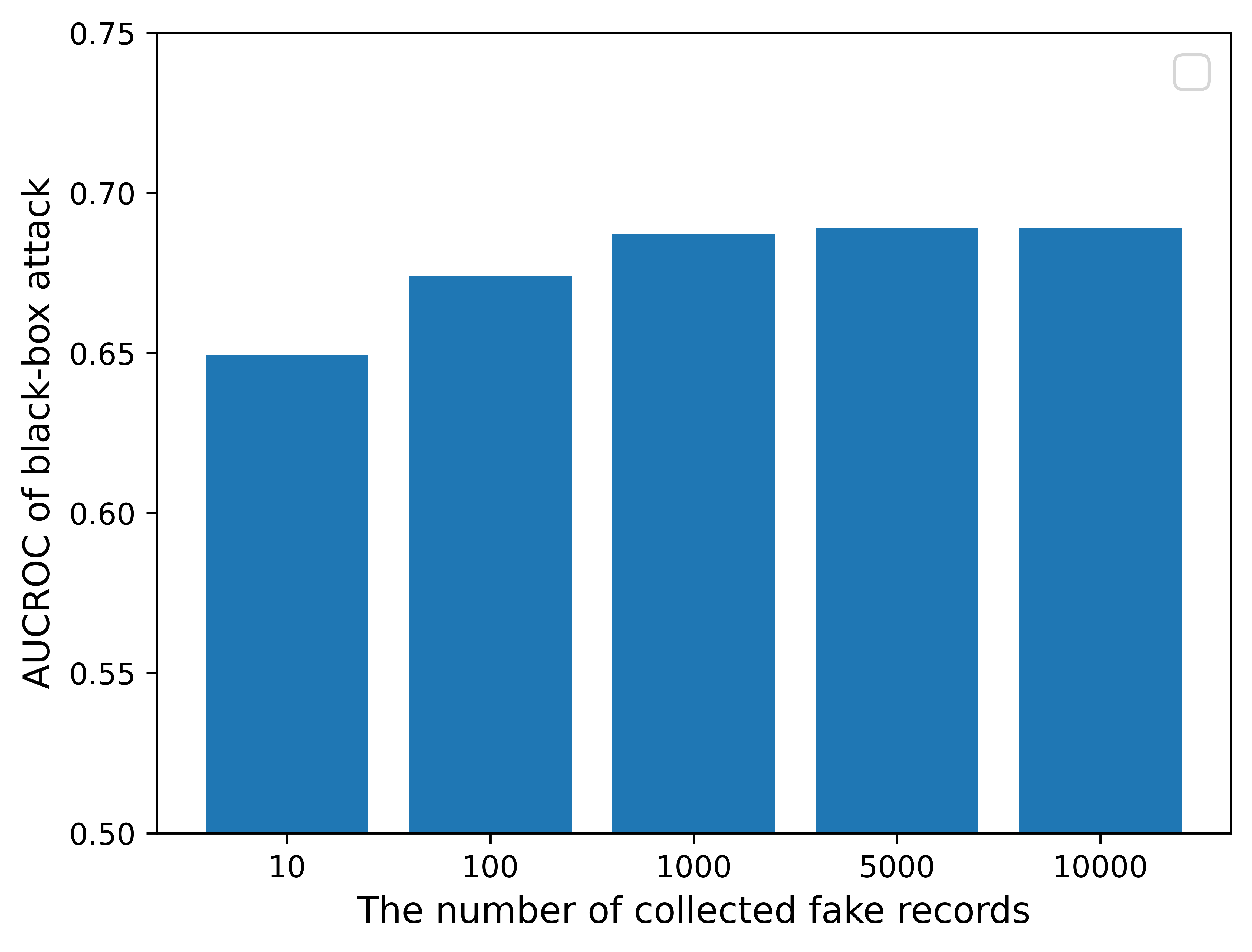}\label{fig4-c}} \hfill
    \subfloat[\texttt{King}]{\includegraphics[width=0.48\columnwidth,trim={0.7cm 0 0 0.28cm},clip]{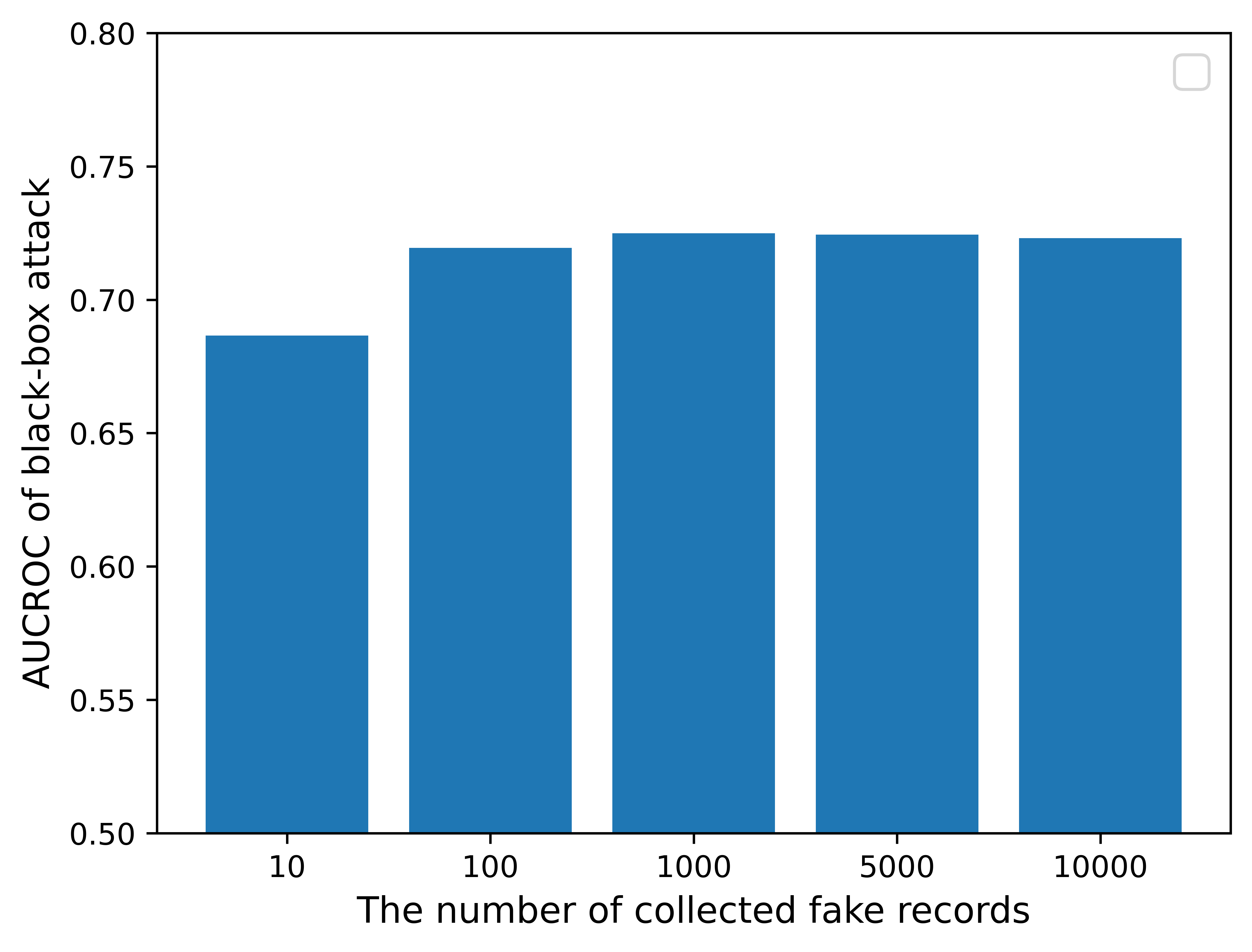}\label{fig1-a}} \\

    \subfloat[\texttt{Adult}]{\includegraphics[width=0.48\columnwidth,trim={0.7cm 0 0 0},clip]{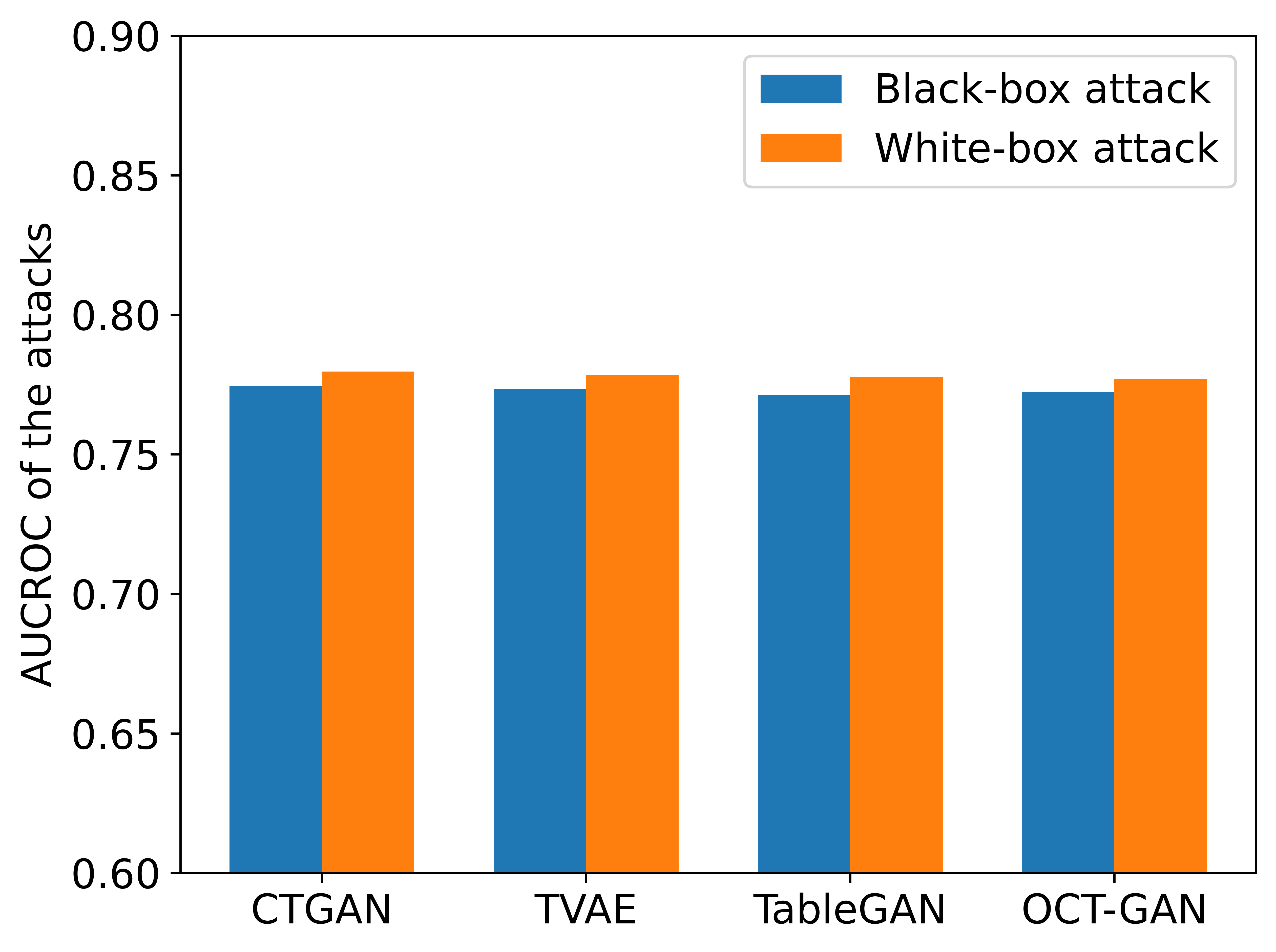} \label{fig5-a}}\hfill
    \subfloat[\texttt{Alphabank}]{\includegraphics[width=0.48\columnwidth,trim={0.7cm 0 0 0.276cm},clip]{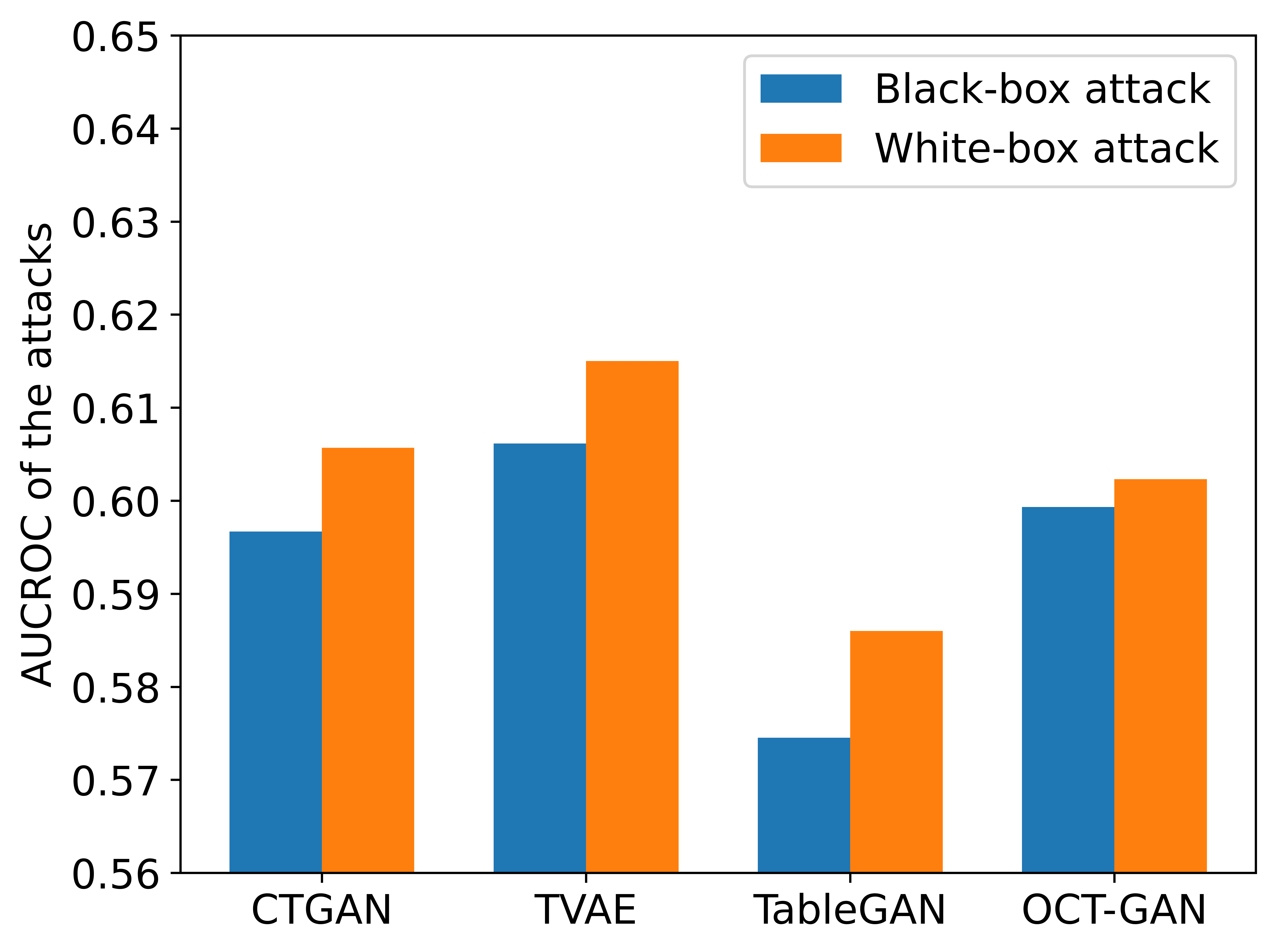}\label{fig1-d}} \hfill
    \subfloat[\texttt{Surgical}]{\includegraphics[width=0.48\columnwidth,trim={0.7cm 0 0 0},clip]{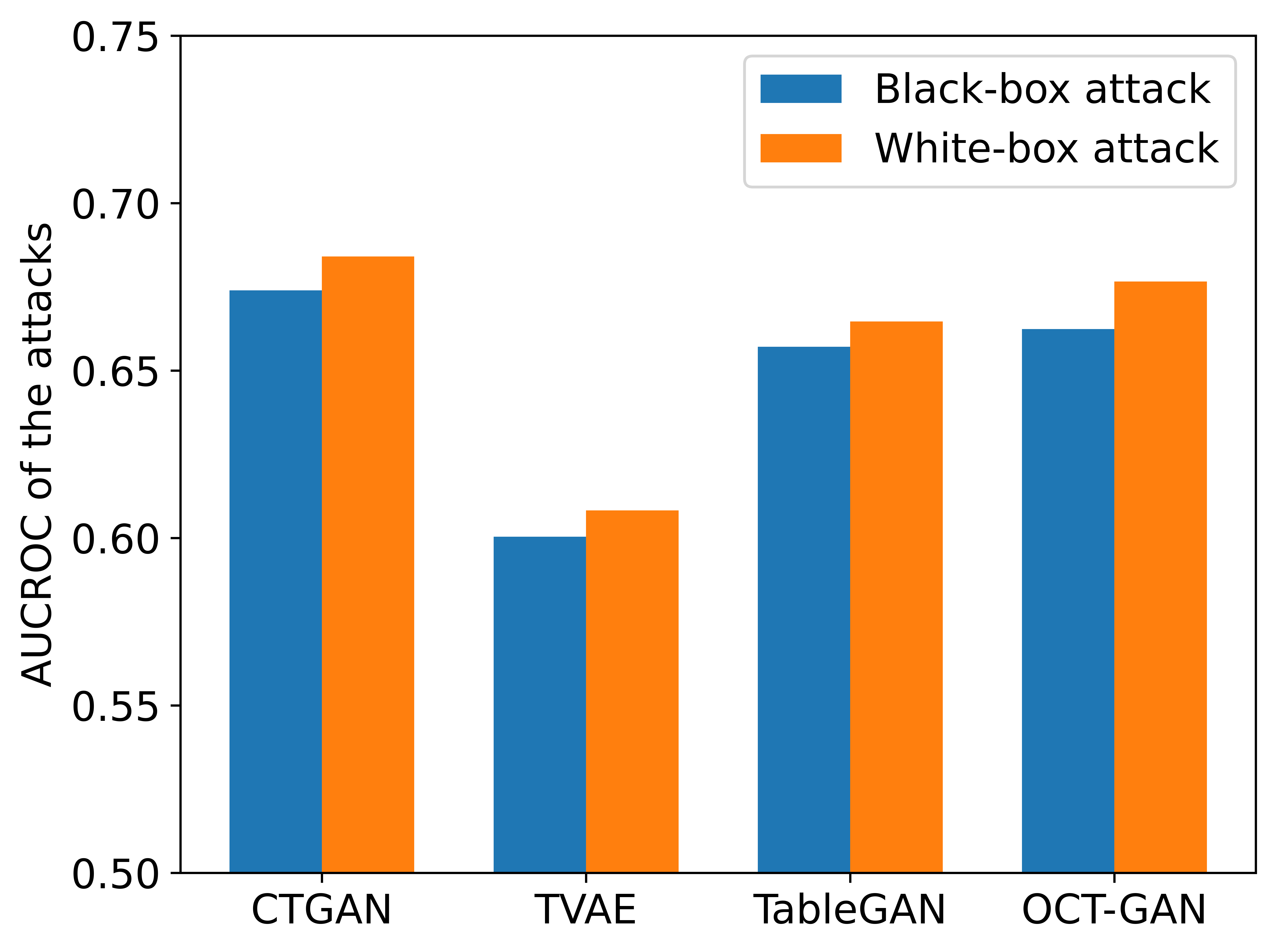}   \label{fig5-c}} \hfill
    \subfloat[\texttt{King}]{\includegraphics[width=0.48\columnwidth,trim={0.7cm 0 0 0.26cm},clip]{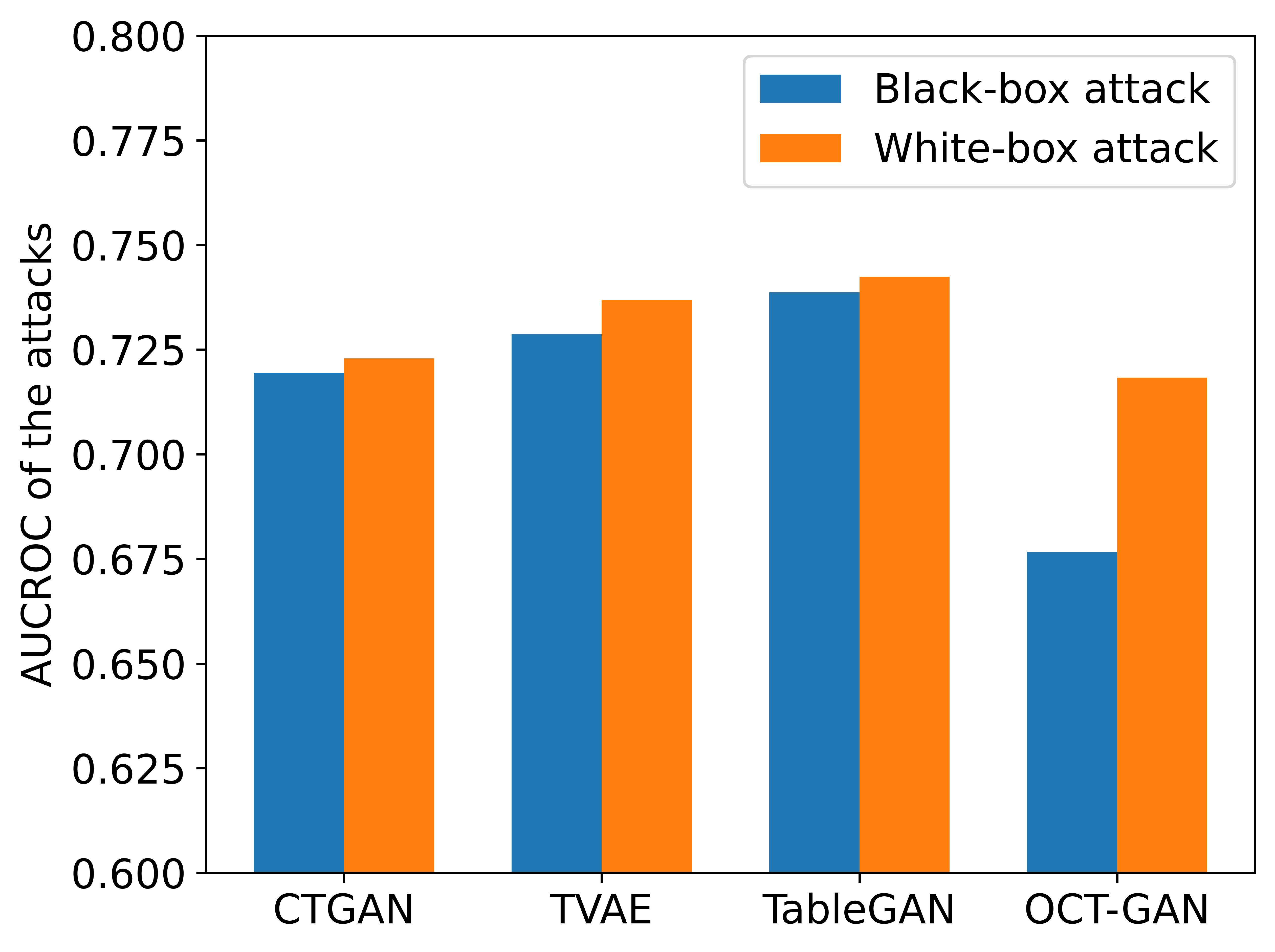}\label{fig1-c}}\\
    
    \caption{(top) The AUROC scores of the black-box attack w.r.t. the number of fake records collected by attackers on \texttt{CTGAN}. The more samples attackers collect, the more successful the attack becomes. After attackers collect more than 100 samples, AUROC scores start to be satiated. (bottom) Comparison between AUROC scores of the black-box attack and the white-box attack for all victim models with 100 generated samples. Victim models are more vulnerable to the white-box attack.}
    \label{fig:fig1}
\end{figure*}

\subsection{Datasets}
We first introduce the following 4 real-world tabular datasets: i) \texttt{Adult} \cite{adult} consists of demographic information to predict whether each person's income exceeds \$50K per year based on census data, and the size of the dataset is 32K records. ii) \texttt{Alphabank} \cite{alphabank} contains 30K records. It is to predict the success of bank telemarketing based on personal financial information. iii) \texttt{King} \cite{king} is a regression dataset to predict house sale prices, which consists of 25K records. iv) \texttt{Surgical} \cite{surgical} is a binary classification dataset, which contains surgical patient information. The dataset contains 13K records.  

\subsection{Victim Generative Models}
As attackers are more likely to target well-trained generative models in practice ~\cite{Chen_2020}, we consider \texttt{CTGAN}, \texttt{TVAE}, \texttt{TableGAN} and \texttt{OCT-GAN} as our victim models, which all show high synthesis quality.

\subsection{Evaluation Method and Metrics}

\begin{table}
\small
    \centering
    \caption{Macro F1 scores ($R^2$ score for \texttt{King}) of victim models with 100 generation samples (1000 for \texttt{King}). All victim models shows reasonable generation quality.}

    \begin{tabular}{c|cccccccccc}
    \specialrule{1pt}{1pt}{1pt}
    \setlength{\tabcolsep}{1pt}
     Dataset  & \texttt{Identity} & \texttt{CTGAN} & \texttt{TVAE} & \texttt{TableGAN} & \texttt{OCT-GAN} & \\ \hline
     \texttt{Adult}   & 0.778 & 0.562 & 0.665 & 0.536 & 0.546 & \\
     \texttt{Alphabank}   & 0.510 & 0.501 & 0.490 & 0.488 & 0.500 & \\
     \texttt{Surgical}   & 0.752 & 0.678 & 0.597 & 0.578 & 0.659 &  \\ 
     \texttt{King}   & 0.421 & 0.250 & 0.444 & 0.357 & 0.293 & \\ \hline

    \end{tabular}
    \label{tab:my_label}
 \end{table}

To evaluate the generation quality, we i) train auxiliary machine learning algorithms, i.e., AdaBoost \cite{adaboost}, DecisionTree \cite{decisiontree}, and Multi-layer Perceptron \cite{mlp}, with synthesized records, and ii) test them on a binary classification or a regression task using test records, which we repeat 10 times with different seeds. We use the following three metrics: i) Macro F1 (resp., $R^2$) score on a binary classification (resp., a regression) task, ii) the average of the minimum column-wise Euclidean distance between real and fake records, and iii) the average of column-wise Earth Mover's Distance (EMD) for the dissimilarity between real and fake records distributions. We present the generation quality of victim models in Table~\ref{tab:my_label}. \texttt{Identity} is a baseline where we perform sampling with replacement from the training set, which means a model whose score is close to \texttt{Identity} has a good generation quality. 

To evaluate attack success scores, we measure the area under the Receiver Operating Characteristic curve (AUROC), one of the most important metrics for binary classification. The most successful the membership inference attack, the close to 1 the AUROC score. If the inference is the same as a random classifier, AUROC is 0.5.


\section{Empirical Study Results}
We summarize our empirical study analysis results for the black-box and the white-box attacks. Source codes and data are available at \url{https://github.com/JayoungKim408/MIA}.
\begin{table*}
\small
\centering
\setlength{\tabcolsep}{3pt}
\caption{Pearson correlation coefficient between the AUROC score of the black-box attack and various generation quality metrics}
\label{tab:correlation}
\begin{tabular}{ccccccccccccccccccccccccc}
\specialrule{1pt}{1pt}{1pt}
 &  & \multicolumn{5}{c}{\texttt{Alphabank}} &  & \multicolumn{5}{c}{\texttt{Adult}} &  & \multicolumn{5}{c}{\texttt{Surgical}} &  & \multicolumn{5}{c}{\texttt{King}} \\ \cline{3-7} \cline{9-13} \cline{15-19} \cline{21-25} 
 &  & \multirow{2}{*}{Macro F1} &  & Euclidean &  & \multirow{2}{*}{EMD} &  & \multirow{2}{*}{Macro F1} &  & Euclidean &  & \multirow{2}{*}{EMD} &  & \multirow{2}{*}{Macro F1} &  & Euclidean &  & \multirow{2}{*}{EMD} &  & \multirow{2}{*}{R2} &  & Euclidean &  & \multirow{2}{*}{EMD} \\
 &  &  &  & distance &  &  &  &  &  & distance &  &  &  &  &  & distance &  &  &  &  &  & distance &  &  \\ \hline
\texttt{CTGAN} &  & 0.927 &  & -0.88 &  & -0.982 &  & 0.998 &  & 0.868 &  & -0.945 &  & 0.980 &  & 0.740 &  & -0.971 &  & 0.891 &  & -0.504 &  & -0.968 \\ 
\texttt{TVAE} &  & 0.849 &  & 0.093 &  & -0.99 &  & 0.883 &  & 0.084 &  & 0.694 &  & 0.995 &  & -0.972 &  & -0.922 &  & 0.869 &  & 0.552 &  & -0.941 \\ 
\texttt{TableGAN} &  & 0.649 &  & 0.888 &  & -0.975 &  & 0.891 &  & 0.959 &  & -0.974 &  & 0.942 &  & -0.924 &  & -0.998 &  & 0.943 &  & -0.810 &  & 0.339 \\ 
\texttt{OCT-GAN} &  & 0.990 &  & 0.873 &  & -0.868 &  & 0.998 &  & -0.933 &  & -0.987 &  & 0.989 &  & -0.990 &  & -0.993 &  & 0.977 &  & 0.339 &  & 0.316 \\ \hline
\end{tabular}
\end{table*}

\subsection{Black-box Attack}
We analyze experimental results on the black-box attack in twofold: i) how much the number of fake records collected by attackers influences the success of the attack and ii) how much the generation quality of victim models is correlated to the success of the attack. Fig.~\ref{fig:fig1} (top) shows the AUROC scores of the black-box attack with respect the number of fake records collected by attackers on \texttt{CTGAN}. There is a clear tendency for AUROC scores to increase with the larger number of collected fake records. As black-box attackers can only access fake records generated by victim models, a larger collection allows membership inference with more knowledge about the victims. Interestingly, the AUROC scores get flattened after more than 100 samples are collected, which means there exists a threshold that attackers may want to consider for practical attack efficiency. 
Table~\ref{tab:correlation} shows Pearson correlation coefficient between the generation quality and the AUROC score of the black-box attack. For Macro F1, larger values are preferred in terms of the generation quality, while for Euclidean distance and EMD, smaller ones are. Thus, a positive (resp., negative) correlation for Macro F1 (resp., for Euclidean distance and EMD) to AUROC indicates that victim models of better generation quality tend to be more vulnerable to the black-box attack. However, as shown in our experimental result, EMD shows a more reliable tendency than Euclidean distance. For example, Euclidean distance has all positive correlations except \texttt{CTGAN} for \texttt{Alphabank}, which happens by the corner case of the metric where most fake records are close to few real records. Notably, in our cases, EMD explains the correlation better than Euclidean distance, which is widely used to evaluate the generation quality in research on attacks~\cite{Chen_2020, hayes2018logan, ChoquetteChoo2020LabelOnlyMI}. 



\subsection{White-box Attack} \label{wb}
As mentioned earlier, the AUROC scores of the black-box attack are satiated with more than 100 fake records collected. For fairness, therefore, we conduct experiments on the white-box attack with 100 collected samples. Note that white-box attackers can generate fake records by themselves from victim models by feeding the noisy vector $\bm{z}$. We let victim models generate 100 fake records for each attack and evaluate AUROC on them. Fig.~\ref{fig:fig1} (bottom) shows a comparison between the AUROC scores of the black and white-box attack on all victim models. In all cases, the white-box attack has higher AUROC scores than the black-box attack, which is intuitive as white-box attackers have full control over victim models. 




\subsection{Defense with Differential Privacy}
We conduct additional experiments on two DP algorithms, DP-SGD and DP-GAN. Due to the scarcity of study about training specific neural network designs with DP algorithms, e.g., moving of \texttt{TableGAN} is not studied, we target only on \texttt{CTGAN}, which consists of trainable network designs by DP algorithms that are currently available. However, as computing the gradient penalty of the WGAN-GP loss~\cite{wgangp} of \texttt{CTGAN} is challenging with DP algorithms, we change it to the Wasserstein loss~\cite{wgan}. Accordingly, we retrain and evaluate the modified \texttt{CTGAN} without DP algorithms as well for an accurate comparison. We repeat each experiment more than 100 times to average out the effect of the randomness in the DP algorithms. 

\begin{figure*}
    \centering
    \captionsetup[subfloat]{justification=centering}
    \subfloat[Macro F1 \\ (trained without DP = 0.60)]{\includegraphics[width=0.48\columnwidth,trim={0.75cm 0 0 0.28cm},clip]{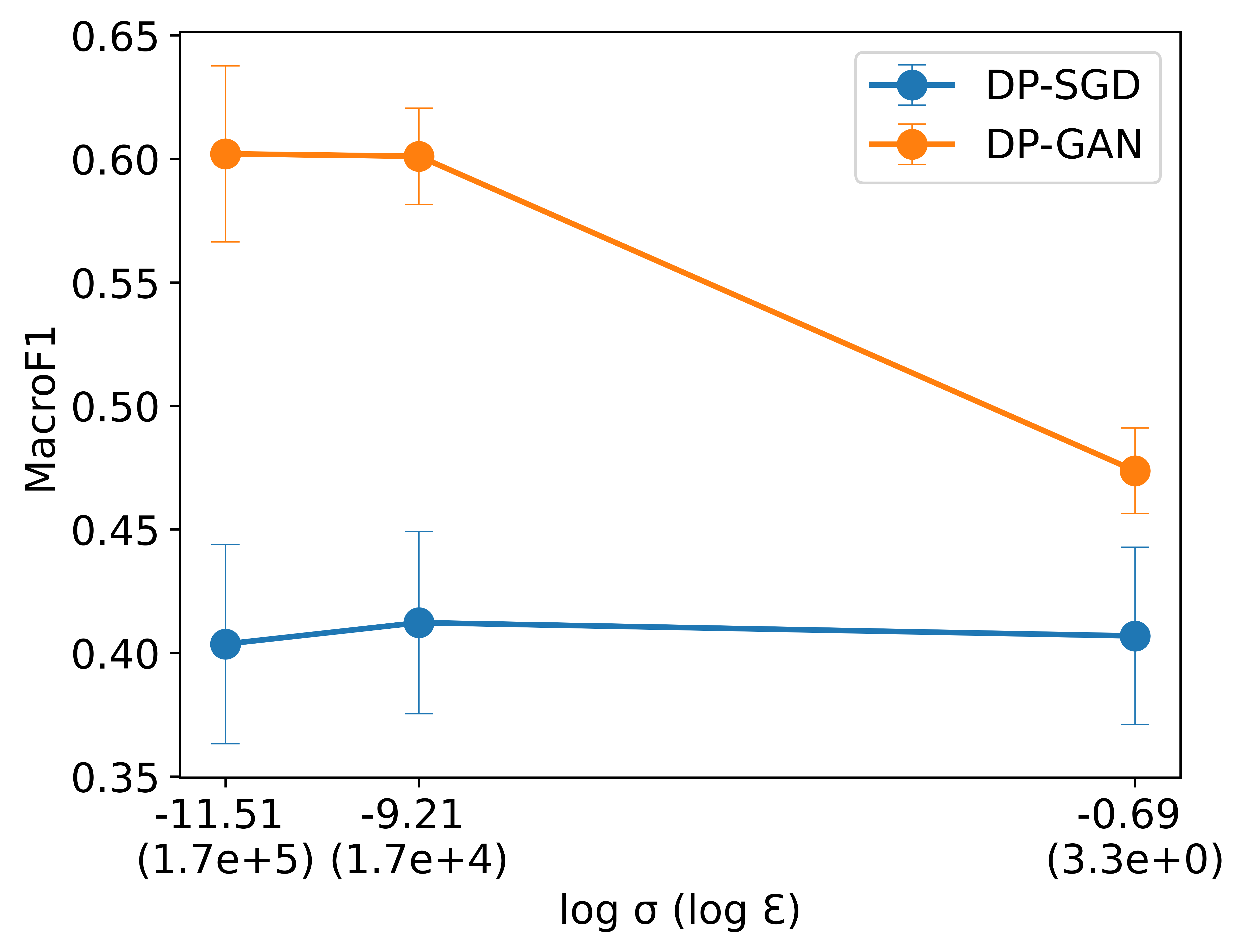}}\hfill
    \subfloat[Euclidean Distance \\(trained without DP = 0.33)]{\includegraphics[width=0.48\columnwidth,trim={0.75cm 0 0 0.25cm},clip]{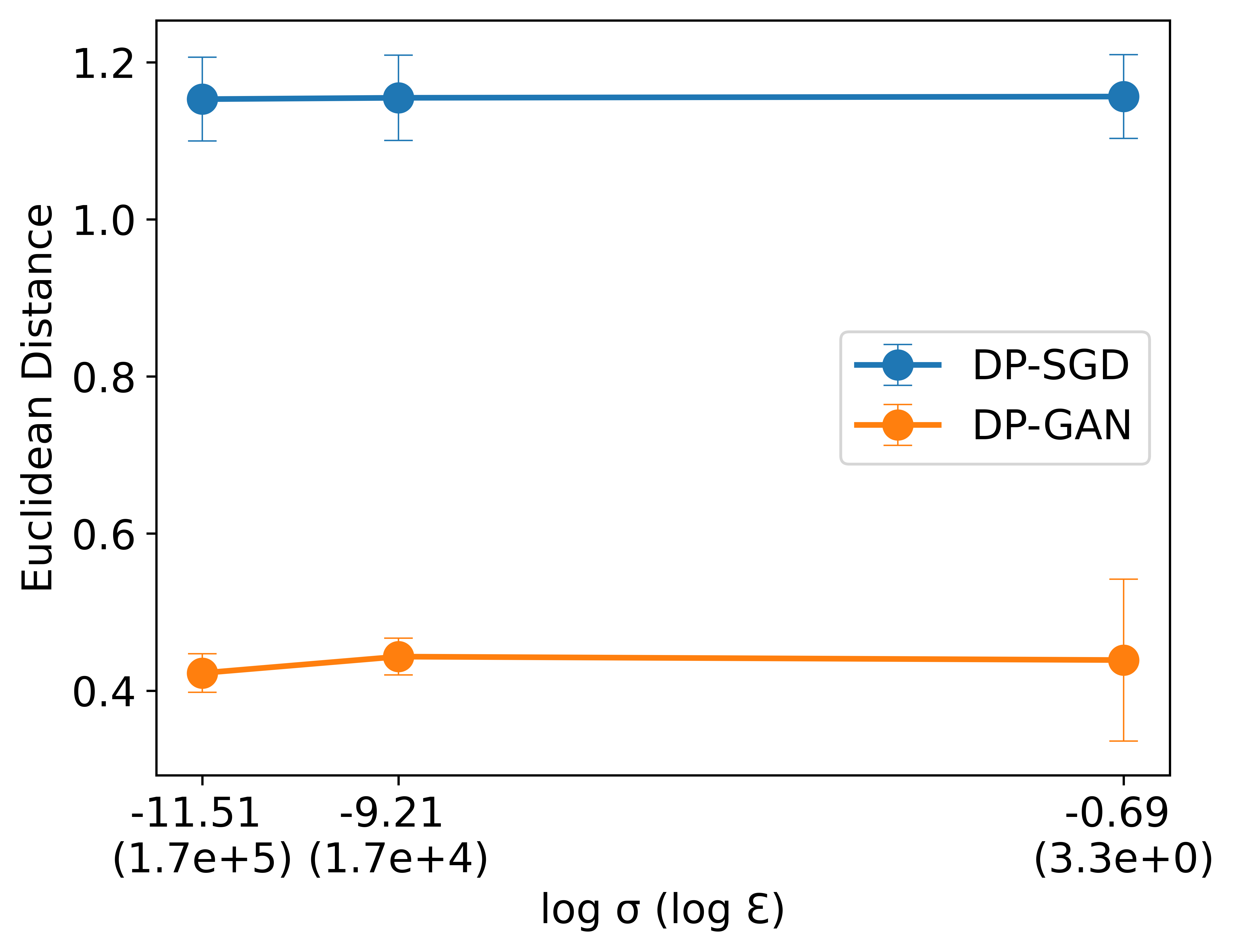}} \hfill 
    \subfloat[AUROC of the black-box attack \\ (trained without DP = 0.77)]{ \includegraphics[width=0.48\columnwidth,trim={0.75cm 0 0 0.25cm},clip]{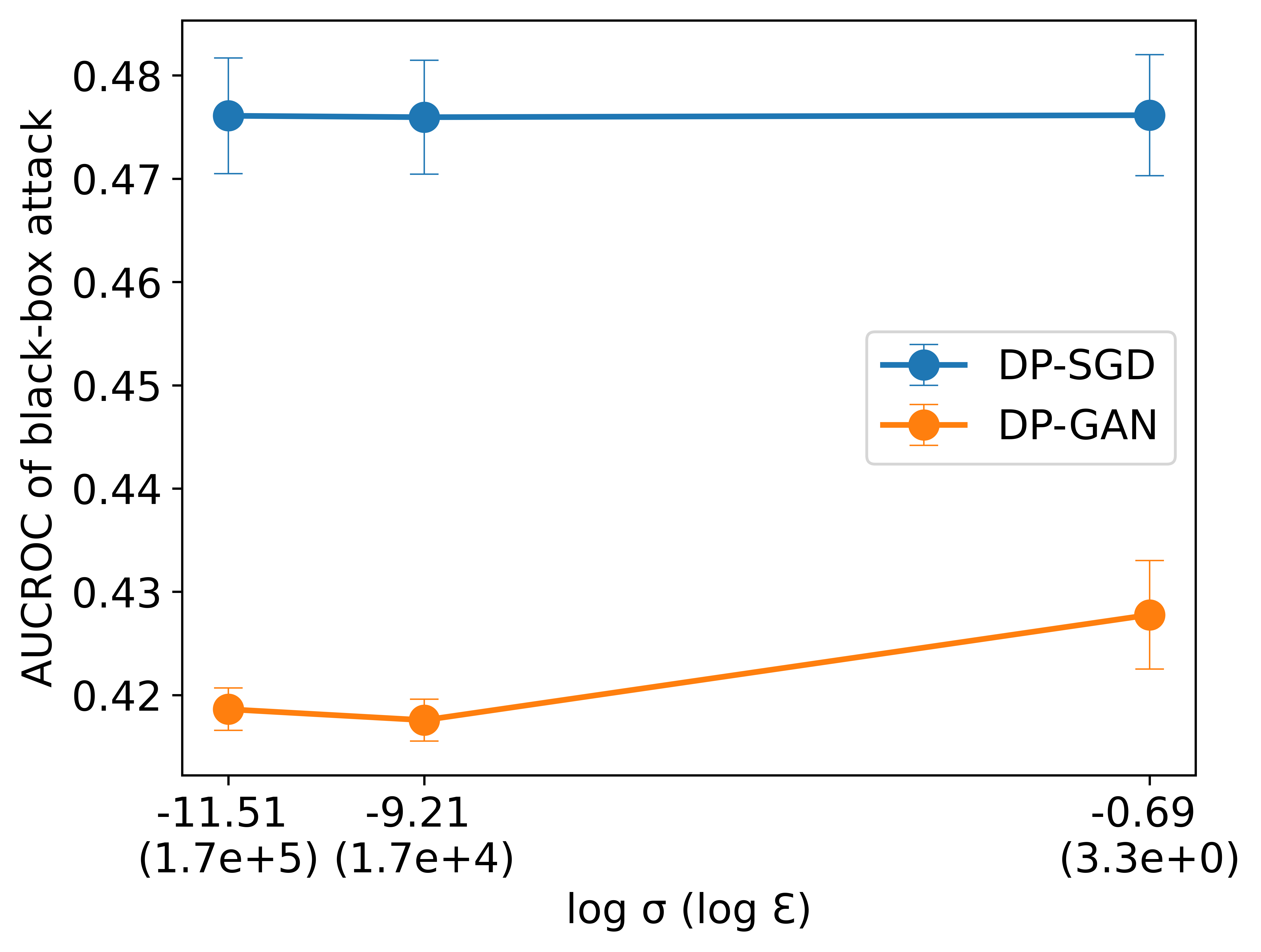}}\hfill
    \subfloat[AUROC of the white-box attack \\ (trained without DP = 0.79)]{\includegraphics[width=0.48\columnwidth,trim={0.75cm 0 0 0.25cm},clip]{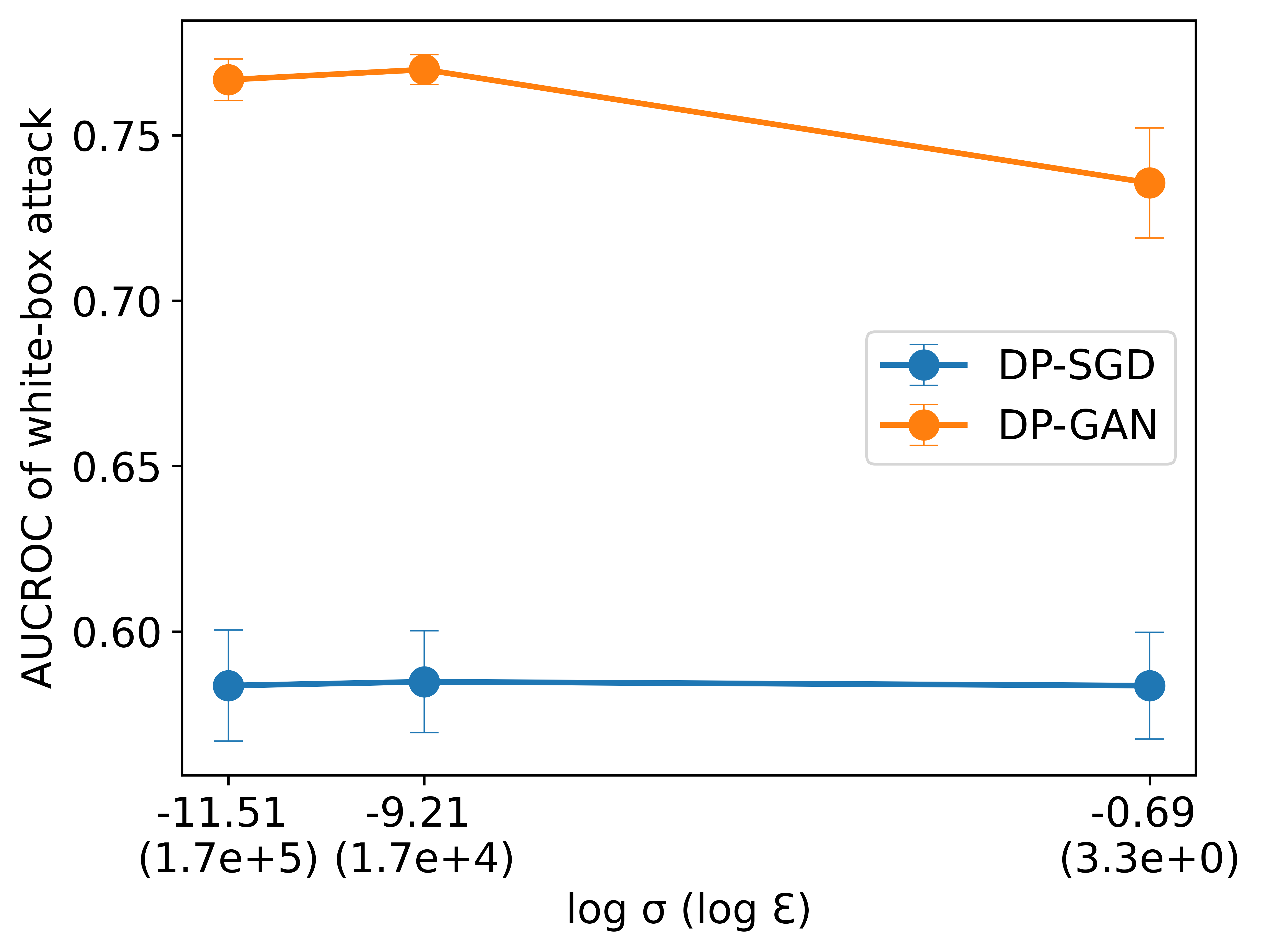}}
            
    \caption{Macro F1, Euclidean distance, and the AUROC scores of the black-box and the white-box attacks on \texttt{CTGAN} for $\texttt{Adult}$ trained with DP-SGD and DP-GAN w.r.t. {$\sigma=\{1e-05, 1e-04, 0.5\}$}, tested on 100 fake record generations. Values in the parentheses under $\sigma$ on X-axis mean $\epsilon$. Values in parentheses contain the results of \texttt{CTGAN} without DP. For DP-GAN, we clip weights to be 1.0. Other datasets show the same pattern, which we refer to Figs.~\ref{fig:fig6}, \ref{fig:fig7} and \ref{fig:fig8}.}
    \label{fig:fig3}
\end{figure*}

Fig.~\ref{fig:fig3} summarizes the comparisons of Macro F1, Euclidean distance, and AUROC of the black and white-box attack on \texttt{CTGAN} for \texttt{Adult} with respect to $\sigma = \{1e-05, 1e-04, 0.5\}$. $\sigma$ controls how much noise to be added to the gradients; thus, there is a trade-off between privacy and model accuracy. In Fig.~\ref{fig:fig3} (a) and (b), compared to training without DP algorithms, training with DP algorithms deteriorates the generation quality. However, DP-GAN shows the generation quality which is better than DP-SGD or even similar to models without DP at lower $\sigma$. Also, DP-GAN worsens the generation quality as $\sigma$ increases, while there is no such pattern for DP-SGD. This is because DP-GAN is specially designed for training GANs with theoretical correctness~\cite{dpgan}. 

In Fig.~\ref{fig:fig3} (c) and (d), models trained with DP algorithms show lower AUROC scores both on the black and white-box attack than models without DP algorithms, as expected. In Fig.~\ref{fig:fig3} (c), DP-GAN shows better robustness to the black-box attack, albeit both show AUROC scores less than 0.5, which means incorrect inference towards membership. Interestingly, the white-box attackers can still infer membership correctly; in Fig.~\ref{fig:fig3} (d), DP-SGD shows better robustness to the white-box attack than DP-GAN; as the white-box attackers can access victim models, DP-GAN, which less deteriorates victims, shows higher vulnerability to attacks. 


\section{Conclusions}
In this paper, we investigated under which conditions tabular data synthesis models are vulnerable to two MIA paradigms, the black-box and the white-box attack, and how defense algorithms can protect them. We conducted experiments with 4 tabular data synthesis models and 4 real-world tabular datasets. 

As expected, the experimental results show that i) as attackers are more knowledgeable about victim models, i.e., when they collect more fake records in the black-box attack or can access the models in the white-box attack, and ii) as models synthesize better representative samples, models are more vulnerable. iii) When models are trained with DP algorithms, models are more robust on the attacks. We also reported detailed findings in Sec. 1.

As shown in experimental results of the defense with two DP algorithms, both have pros and cons; DP-SGD largely ruins the generation quality of victim models while protecting them against the white-box attack; DP-GAN preserves the generation quality of victim models while exposing them more to the white-box attack. This implies that one may decide how to defend tabular data synthesis models based on his/her purpose. For instance, one may sacrifice high-quality generation for enhanced privacy guarantee by adopting DP-SGD. This also motivates future work of combining two different approaches from DP-SGD and DP-GAN.

Our paper delivered several empirical messages and findings that are hard to know without conducting all those experiments. We think our paper can enlighten and guide others worried about privacy concerns regarding their fake tabular data.

\begin{acks}
Noseong Park is the corresponding author. This work was supported by the Institute of Information \& Communications Technology Planning \& Evaluation (IITP) grant funded by the Korean government (MSIT) (10\% from No. 2020-0-01361, Artificial Intelligence Graduate School Program at (Yonsei University and 90\% from No. 2021-0-00231, Development of Approximate DBMS Query Technology to Facilitate Fast Query Processing for Exploratory Data Analysis)). The work by Sushil Jajodia was supported by the U.S. Office of Naval Research grants N00014-20-1-2407 and N00014-18-1-2670, and by the U.S. National Science Foundation grant CNS-1822094.
\end{acks}


\clearpage
\bibliographystyle{ACM-Reference-Format}
\bibliography{arxiv_full}


\begin{thebibliography}{29}


\ifx \showCODEN    \undefined \def \showCODEN     #1{\unskip}     \fi
\ifx \showDOI      \undefined \def \showDOI       #1{#1}\fi
\ifx \showISBNx    \undefined \def \showISBNx     #1{\unskip}     \fi
\ifx \showISBNxiii \undefined \def \showISBNxiii  #1{\unskip}     \fi
\ifx \showISSN     \undefined \def \showISSN      #1{\unskip}     \fi
\ifx \showLCCN     \undefined \def \showLCCN      #1{\unskip}     \fi
\ifx \shownote     \undefined \def \shownote      #1{#1}          \fi
\ifx \showarticletitle \undefined \def \showarticletitle #1{#1}   \fi
\ifx \showURL      \undefined \def \showURL       {\relax}        \fi
\providecommand\bibfield[2]{#2}
\providecommand\bibinfo[2]{#2}
\providecommand\natexlab[1]{#1}
\providecommand\showeprint[2][]{arXiv:#2}

\bibitem[\protect\citeauthoryear{Abadi, Chu, Goodfellow, McMahan, Mironov,
  Talwar, and Zhang}{Abadi et~al\mbox{.}}{2016}]%
        {Abadi_2016}
\bibfield{author}{\bibinfo{person}{Martin Abadi}, \bibinfo{person}{Andy Chu},
  \bibinfo{person}{Ian Goodfellow}, \bibinfo{person}{H.~Brendan McMahan},
  \bibinfo{person}{Ilya Mironov}, \bibinfo{person}{Kunal Talwar}, {and}
  \bibinfo{person}{Li Zhang}.} \bibinfo{year}{2016}\natexlab{}.
\newblock \showarticletitle{Deep Learning with Differential Privacy}. In
  \bibinfo{booktitle}{\emph{Proceedings of the 2016 ACM SIGSAC Conference on
  Computer and Communications Security}} (Vienna, Austria)
  \emph{(\bibinfo{series}{CCS '16})}. \bibinfo{publisher}{Association for
  Computing Machinery}, \bibinfo{address}{New York, NY, USA},
  \bibinfo{pages}{308–318}.
\newblock
\showISBNx{9781450341394}
\urldef\tempurl%
\url{https://doi.org/10.1145/2976749.2978318}
\showDOI{\tempurl}


\bibitem[\protect\citeauthoryear{Arjovsky, Chintala, and Bottou}{Arjovsky
  et~al\mbox{.}}{2017}]%
        {wgan}
\bibfield{author}{\bibinfo{person}{Martin Arjovsky}, \bibinfo{person}{Soumith
  Chintala}, {and} \bibinfo{person}{Léon Bottou}.}
  \bibinfo{year}{2017}\natexlab{}.
\newblock \bibinfo{title}{Wasserstein GAN}.
\newblock
\newblock
\urldef\tempurl%
\url{http://arxiv.org/abs/1701.07875}
\showURL{%
\tempurl}
\newblock
\shownote{cite arxiv:1701.07875.}


\bibitem[\protect\citeauthoryear{Bayardo and Agrawal}{Bayardo and
  Agrawal}{2005}]%
        {anonymization}
\bibfield{author}{\bibinfo{person}{Roberto~J. Bayardo} {and}
  \bibinfo{person}{Rakesh Agrawal}.} \bibinfo{year}{2005}\natexlab{}.
\newblock \showarticletitle{Data Privacy through Optimal K-Anonymization}. In
  \bibinfo{booktitle}{\emph{Proceedings of the 21st International Conference on
  Data Engineering}} \emph{(\bibinfo{series}{ICDE '05})}.
  \bibinfo{publisher}{IEEE Computer Society}, \bibinfo{address}{USA},
  \bibinfo{pages}{217–228}.
\newblock
\showISBNx{0769522858}
\urldef\tempurl%
\url{https://doi.org/10.1109/ICDE.2005.42}
\showDOI{\tempurl}


\bibitem[\protect\citeauthoryear{Beaulieu-Jones, Wu, Williams, Lee, Bhavnani,
  Byrd, and Greene}{Beaulieu-Jones et~al\mbox{.}}{2019}]%
        {beaulieu2019privacy}
\bibfield{author}{\bibinfo{person}{Brett Beaulieu-Jones},
  \bibinfo{person}{Zhiwei Wu}, \bibinfo{person}{Chris Williams},
  \bibinfo{person}{Ran Lee}, \bibinfo{person}{Sanjeev Bhavnani},
  \bibinfo{person}{James Byrd}, {and} \bibinfo{person}{Casey Greene}.}
  \bibinfo{year}{2019}\natexlab{}.
\newblock \showarticletitle{Privacy-Preserving Generative Deep Neural Networks
  Support Clinical Data Sharing}.
\newblock \bibinfo{journal}{\emph{Circulation: Cardiovascular Quality and
  Outcomes}}  \bibinfo{volume}{12} (\bibinfo{date}{07} \bibinfo{year}{2019}).
\newblock
\urldef\tempurl%
\url{https://doi.org/10.1161/CIRCOUTCOMES.118.005122}
\showDOI{\tempurl}


\bibitem[\protect\citeauthoryear{Bishop}{Bishop}{2006}]%
        {mlp}
\bibfield{author}{\bibinfo{person}{Christopher~M. Bishop}.}
  \bibinfo{year}{2006}\natexlab{}.
\newblock \bibinfo{booktitle}{\emph{Pattern Recognition and Machine Learning
  (Information Science and Statistics)}}.
\newblock \bibinfo{publisher}{Springer-Verlag}, \bibinfo{address}{Berlin,
  Heidelberg}.
\newblock
\showISBNx{0387310738}


\bibitem[\protect\citeauthoryear{Chen, Yu, Zhang, and Fritz}{Chen
  et~al\mbox{.}}{2020}]%
        {Chen_2020}
\bibfield{author}{\bibinfo{person}{Dingfan Chen}, \bibinfo{person}{Ning Yu},
  \bibinfo{person}{Yang Zhang}, {and} \bibinfo{person}{Mario Fritz}.}
  \bibinfo{year}{2020}\natexlab{}.
\newblock \showarticletitle{GAN-Leaks: A Taxonomy of Membership Inference
  Attacks against Generative Models}. In \bibinfo{booktitle}{\emph{Proceedings
  of the 2020 ACM SIGSAC Conference on Computer and Communications Security}}
  (Virtual Event, USA) \emph{(\bibinfo{series}{CCS '20})}.
  \bibinfo{publisher}{Association for Computing Machinery},
  \bibinfo{address}{New York, NY, USA}, \bibinfo{pages}{343–362}.
\newblock
\showISBNx{9781450370899}
\urldef\tempurl%
\url{https://doi.org/10.1145/3372297.3417238}
\showDOI{\tempurl}


\bibitem[\protect\citeauthoryear{Chen, Xiang, Xue, Li, Borisov, Kaarfar, and
  Zhu}{Chen et~al\mbox{.}}{2018}]%
        {chen2018differentially}
\bibfield{author}{\bibinfo{person}{Qingrong Chen}, \bibinfo{person}{Chong
  Xiang}, \bibinfo{person}{Minhui Xue}, \bibinfo{person}{Bo Li},
  \bibinfo{person}{Nikita Borisov}, \bibinfo{person}{Dali Kaarfar}, {and}
  \bibinfo{person}{Haojin Zhu}.} \bibinfo{year}{2018}\natexlab{}.
\newblock \bibinfo{title}{Differentially Private Data Generative Models}.
\newblock
\newblock
\showeprint[arxiv]{1812.02274}~[cs.CR]


\bibitem[\protect\citeauthoryear{Choquette-Choo, Tramer, Carlini, and
  Papernot}{Choquette-Choo et~al\mbox{.}}{2021}]%
        {ChoquetteChoo2020LabelOnlyMI}
\bibfield{author}{\bibinfo{person}{Christopher~A. Choquette-Choo},
  \bibinfo{person}{Florian Tramer}, \bibinfo{person}{Nicholas Carlini}, {and}
  \bibinfo{person}{Nicolas Papernot}.} \bibinfo{year}{2021}\natexlab{}.
\newblock \bibinfo{title}{Label-Only Membership Inference Attacks}.
\newblock , \bibinfo{numpages}{1964--1974}~pages.
\newblock
\urldef\tempurl%
\url{https://proceedings.mlr.press/v139/choquette-choo21a.html}
\showURL{%
\tempurl}


\bibitem[\protect\citeauthoryear{County.}{County.}{2016}]%
        {king}
\bibfield{author}{\bibinfo{person}{King County.}}
  \bibinfo{year}{2016}\natexlab{}.
\newblock \bibinfo{title}{House Sales in King County, USA}.
\newblock
\newblock


\bibitem[\protect\citeauthoryear{Duda, Hart, and Stork}{Duda
  et~al\mbox{.}}{2001}]%
        {DudaHartStork01}
\bibfield{author}{\bibinfo{person}{Richard~O. Duda}, \bibinfo{person}{Peter~E.
  Hart}, {and} \bibinfo{person}{David~G. Stork}.}
  \bibinfo{year}{2001}\natexlab{}.
\newblock \bibinfo{booktitle}{\emph{Pattern Classification}
  (\bibinfo{edition}{2} ed.)}.
\newblock \bibinfo{publisher}{Wiley}, \bibinfo{address}{New York}.
\newblock
\showISBNx{978-0-471-05669-0}


\bibitem[\protect\citeauthoryear{Dwork}{Dwork}{2006}]%
        {Dwork}
\bibfield{author}{\bibinfo{person}{Cynthia Dwork}.}
  \bibinfo{year}{2006}\natexlab{}.
\newblock \showarticletitle{{Differential Privacy}}. In
  \bibinfo{booktitle}{\emph{Automata, Languages and Programming}},
  \bibfield{editor}{\bibinfo{person}{Michele Bugliesi}, \bibinfo{person}{Bart
  Preneel}, \bibinfo{person}{Vladimiro Sassone}, {and} \bibinfo{person}{Ingo
  Wegener}} (Eds.). \bibinfo{publisher}{Springer Berlin Heidelberg},
  \bibinfo{address}{Berlin, Heidelberg}, \bibinfo{pages}{1--12}.
\newblock
\showISBNx{978-3-540-35908-1}


\bibitem[\protect\citeauthoryear{Dwork, Roth, et~al\mbox{.}}{Dwork
  et~al\mbox{.}}{2014}]%
        {dwork2014algorithmic}
\bibfield{author}{\bibinfo{person}{Cynthia Dwork}, \bibinfo{person}{Aaron
  Roth}, {et~al\mbox{.}}} \bibinfo{year}{2014}\natexlab{}.
\newblock \showarticletitle{The algorithmic foundations of differential
  privacy.}
\newblock \bibinfo{journal}{\emph{Found. Trends Theor. Comput. Sci.}}
  \bibinfo{volume}{9}, \bibinfo{number}{3-4} (\bibinfo{year}{2014}),
  \bibinfo{pages}{211--407}.
\newblock


\bibitem[\protect\citeauthoryear{Gulrajani, Ahmed, Arjovsky, Dumoulin, and
  Courville}{Gulrajani et~al\mbox{.}}{2017}]%
        {wgangp}
\bibfield{author}{\bibinfo{person}{Ishaan Gulrajani}, \bibinfo{person}{Faruk
  Ahmed}, \bibinfo{person}{Martin Arjovsky}, \bibinfo{person}{Vincent
  Dumoulin}, {and} \bibinfo{person}{Aaron Courville}.}
  \bibinfo{year}{2017}\natexlab{}.
\newblock \showarticletitle{Improved Training of Wasserstein GANs}. In
  \bibinfo{booktitle}{\emph{Proceedings of the 31st International Conference on
  Neural Information Processing Systems}} (Long Beach, California, USA)
  \emph{(\bibinfo{series}{NIPS'17})}. \bibinfo{publisher}{Curran Associates
  Inc.}, \bibinfo{address}{Red Hook, NY, USA}, \bibinfo{pages}{5769–5779}.
\newblock
\showISBNx{9781510860964}


\bibitem[\protect\citeauthoryear{Hayes, Melis, Danezis, and Cristofaro}{Hayes
  et~al\mbox{.}}{2018}]%
        {hayes2018logan}
\bibfield{author}{\bibinfo{person}{Jamie Hayes}, \bibinfo{person}{Luca Melis},
  \bibinfo{person}{George Danezis}, {and} \bibinfo{person}{Emiliano~De
  Cristofaro}.} \bibinfo{year}{2018}\natexlab{}.
\newblock \bibinfo{title}{LOGAN: Membership Inference Attacks Against
  Generative Models}.
\newblock
\newblock
\showeprint[arxiv]{1705.07663}~[cs.CR]


\bibitem[\protect\citeauthoryear{Hilprecht, H{\"a}rterich, and
  Bernau}{Hilprecht et~al\mbox{.}}{2019}]%
        {hilprecht2019monte}
\bibfield{author}{\bibinfo{person}{Benjamin Hilprecht}, \bibinfo{person}{Martin
  H{\"a}rterich}, {and} \bibinfo{person}{Daniel Bernau}.}
  \bibinfo{year}{2019}\natexlab{}.
\newblock \showarticletitle{Monte Carlo and Reconstruction Membership Inference
  Attacks against Generative Models.}
\newblock \bibinfo{journal}{\emph{Proc. Priv. Enhancing Technol.}}
  \bibinfo{volume}{2019}, \bibinfo{number}{4} (\bibinfo{year}{2019}),
  \bibinfo{pages}{232--249}.
\newblock


\bibitem[\protect\citeauthoryear{Ishfaq, Hoogi, and Rubin}{Ishfaq
  et~al\mbox{.}}{2018}]%
        {ishfaq2018tvae}
\bibfield{author}{\bibinfo{person}{Haque Ishfaq}, \bibinfo{person}{Assaf
  Hoogi}, {and} \bibinfo{person}{Daniel Rubin}.}
  \bibinfo{year}{2018}\natexlab{}.
\newblock \bibinfo{title}{TVAE: Triplet-Based Variational Autoencoder using
  Metric Learning}.
\newblock
\newblock
\showeprint{1802.04403}


\bibitem[\protect\citeauthoryear{Jordon, Yoon, and Schaar}{Jordon
  et~al\mbox{.}}{2019}]%
        {Jordon2019PATEGANGS}
\bibfield{author}{\bibinfo{person}{James Jordon}, \bibinfo{person}{Jinsung
  Yoon}, {and} \bibinfo{person}{V.~D.~Mihaela Schaar}.}
  \bibinfo{year}{2019}\natexlab{}.
\newblock \bibinfo{title}{PATE-GAN: Generating Synthetic Data with Differential
  Privacy Guarantees}.
\newblock
\newblock


\bibitem[\protect\citeauthoryear{Kim, Jeon, Lee, Hyeong, and Park}{Kim
  et~al\mbox{.}}{2021}]%
        {kim2021octgan}
\bibfield{author}{\bibinfo{person}{Jayoung Kim}, \bibinfo{person}{Jinsung
  Jeon}, \bibinfo{person}{Jaehoon Lee}, \bibinfo{person}{Jihyeon Hyeong}, {and}
  \bibinfo{person}{Noseong Park}.} \bibinfo{year}{2021}\natexlab{}.
\newblock \bibinfo{title}{OCT-GAN: Neural ODE-based Conditional Tabular GANs}.
\newblock
\newblock
\showeprint[arxiv]{2105.14969}~[cs.LG]


\bibitem[\protect\citeauthoryear{Kohavi}{Kohavi}{1996}]%
        {adult}
\bibfield{author}{\bibinfo{person}{Ron Kohavi}.}
  \bibinfo{year}{1996}\natexlab{}.
\newblock \bibinfo{title}{Scaling up the Accuracy of Naive-Bayes Classifiers: A
  Decision-Tree Hybrid}.
\newblock , \bibinfo{numpages}{6}~pages.
\newblock


\bibitem[\protect\citeauthoryear{Moro, Cortez, and Rita}{Moro
  et~al\mbox{.}}{2014}]%
        {alphabank}
\bibfield{author}{\bibinfo{person}{S{\'e}rgio Moro}, \bibinfo{person}{P.
  Cortez}, {and} \bibinfo{person}{P. Rita}.} \bibinfo{year}{2014}\natexlab{}.
\newblock \showarticletitle{A data-driven approach to predict the success of
  bank telemarketing}.
\newblock \bibinfo{journal}{\emph{Decis. Support Syst.}}  \bibinfo{volume}{62}
  (\bibinfo{year}{2014}), \bibinfo{pages}{22--31}.
\newblock


\bibitem[\protect\citeauthoryear{Park, Mohammadi, Gorde, Jajodia, Park, and
  Kim}{Park et~al\mbox{.}}{2018}]%
        {DBLP:journals/corr/abs-1806-03384}
\bibfield{author}{\bibinfo{person}{Noseong Park}, \bibinfo{person}{Mahmoud
  Mohammadi}, \bibinfo{person}{Kshitij Gorde}, \bibinfo{person}{Sushil
  Jajodia}, \bibinfo{person}{Hongkyu Park}, {and} \bibinfo{person}{Youngmin
  Kim}.} \bibinfo{year}{2018}\natexlab{}.
\newblock \showarticletitle{Data Synthesis Based on Generative Adversarial
  Networks}.
\newblock \bibinfo{journal}{\emph{Proc. VLDB Endow.}} \bibinfo{volume}{11},
  \bibinfo{number}{10} (\bibinfo{date}{jun} \bibinfo{year}{2018}),
  \bibinfo{pages}{1071–1083}.
\newblock
\showISSN{2150-8097}
\urldef\tempurl%
\url{https://doi.org/10.14778/3231751.3231757}
\showDOI{\tempurl}


\bibitem[\protect\citeauthoryear{Reiter}{Reiter}{2005}]%
        {decisiontree}
\bibfield{author}{\bibinfo{person}{P.~Jerome Reiter}.}
  \bibinfo{year}{2005}\natexlab{}.
\newblock \showarticletitle{Using CART to Generate Partially Synthetic, Public
  Use Microdata}.
\newblock \bibinfo{journal}{\emph{Journal of Official Statistics}}
  \bibinfo{volume}{21} (\bibinfo{date}{01} \bibinfo{year}{2005}),
  \bibinfo{pages}{441}.
\newblock


\bibitem[\protect\citeauthoryear{Samarati}{Samarati}{2001}]%
        {kanonymity}
\bibfield{author}{\bibinfo{person}{Pierangela Samarati}.}
  \bibinfo{year}{2001}\natexlab{}.
\newblock \showarticletitle{Protecting Respondents' Identities in Microdata
  Release}.
\newblock \bibinfo{journal}{\emph{IEEE Trans. Knowl. Data Eng.}}
  \bibinfo{volume}{13} (\bibinfo{year}{2001}), \bibinfo{pages}{1010--1027}.
\newblock


\bibitem[\protect\citeauthoryear{Schapire}{Schapire}{1999}]%
        {adaboost}
\bibfield{author}{\bibinfo{person}{Robert~E. Schapire}.}
  \bibinfo{year}{1999}\natexlab{}.
\newblock \showarticletitle{A Brief Introduction to Boosting}. In
  \bibinfo{booktitle}{\emph{Proceedings of the 16th International Joint
  Conference on Artificial Intelligence - Volume 2}} (Stockholm, Sweden)
  \emph{(\bibinfo{series}{IJCAI'99})}. \bibinfo{publisher}{Morgan Kaufmann
  Publishers Inc.}, \bibinfo{address}{San Francisco, CA, USA},
  \bibinfo{pages}{1401–1406}.
\newblock


\bibitem[\protect\citeauthoryear{Sessler, Kurz, Saager, and Dalton}{Sessler
  et~al\mbox{.}}{2011}]%
        {surgical}
\bibfield{author}{\bibinfo{person}{Daniel Sessler}, \bibinfo{person}{Andrea
  Kurz}, \bibinfo{person}{Leif Saager}, {and} \bibinfo{person}{Jarrod Dalton}.}
  \bibinfo{year}{2011}\natexlab{}.
\newblock \showarticletitle{Operation Timing and 30-Day Mortality After
  Elective General Surgery}.
\newblock \bibinfo{journal}{\emph{Anesthesia and analgesia}}
  \bibinfo{volume}{113} (\bibinfo{date}{09} \bibinfo{year}{2011}),
  \bibinfo{pages}{1423--8}.
\newblock
\urldef\tempurl%
\url{https://doi.org/10.1213/ANE.0b013e3182315a6d}
\showDOI{\tempurl}


\bibitem[\protect\citeauthoryear{Shokri, Stronati, Song, and Shmatikov}{Shokri
  et~al\mbox{.}}{2017}]%
        {shokri2017membership}
\bibfield{author}{\bibinfo{person}{Reza Shokri}, \bibinfo{person}{Marco
  Stronati}, \bibinfo{person}{Congzheng Song}, {and} \bibinfo{person}{Vitaly
  Shmatikov}.} \bibinfo{year}{2017}\natexlab{}.
\newblock \bibinfo{title}{Membership Inference Attacks against Machine Learning
  Models}.
\newblock
\newblock
\showeprint[arxiv]{1610.05820}~[cs.CR]


\bibitem[\protect\citeauthoryear{Xie, Lin, Wang, Wang, and Zhou}{Xie
  et~al\mbox{.}}{2018}]%
        {dpgan}
\bibfield{author}{\bibinfo{person}{Liyang Xie}, \bibinfo{person}{Kaixiang Lin},
  \bibinfo{person}{Shu Wang}, \bibinfo{person}{Fei Wang}, {and}
  \bibinfo{person}{Jiayu Zhou}.} \bibinfo{year}{2018}\natexlab{}.
\newblock \bibinfo{title}{Differentially Private Generative Adversarial
  Network}.
\newblock
\newblock
\showeprint[arXiv]{1802.06739}
\urldef\tempurl%
\url{http://arxiv.org/abs/1802.06739}
\showURL{%
\tempurl}


\bibitem[\protect\citeauthoryear{Xu, Ren, Lin, and Sun}{Xu
  et~al\mbox{.}}{2018}]%
        {xu2018dpgan}
\bibfield{author}{\bibinfo{person}{Jingjing Xu}, \bibinfo{person}{Xuancheng
  Ren}, \bibinfo{person}{Junyang Lin}, {and} \bibinfo{person}{Xu Sun}.}
  \bibinfo{year}{2018}\natexlab{}.
\newblock \bibinfo{title}{DP-GAN: Diversity-Promoting Generative Adversarial
  Network for Generating Informative and Diversified Text}.
\newblock
\newblock
\showeprint[arxiv]{1802.01345}~[cs.CL]


\bibitem[\protect\citeauthoryear{Xu, Skoularidou, Cuesta-Infante, and
  Veeramachaneni}{Xu et~al\mbox{.}}{2019}]%
        {NIPS2019_8953}
\bibfield{author}{\bibinfo{person}{Lei Xu}, \bibinfo{person}{Maria
  Skoularidou}, \bibinfo{person}{Alfredo Cuesta-Infante}, {and}
  \bibinfo{person}{Kalyan Veeramachaneni}.} \bibinfo{year}{2019}\natexlab{}.
\newblock \showarticletitle{Modeling Tabular Data Using Conditional GAN}. In
  \bibinfo{booktitle}{\emph{Proceedings of the 33rd International Conference on
  Neural Information Processing Systems}}. \bibinfo{publisher}{Curran
  Associates Inc.}, \bibinfo{address}{Red Hook, NY, USA}, Article
  \bibinfo{articleno}{659}, \bibinfo{numpages}{11}~pages.
\newblock


\end{thebibliography}



\clearpage
\appendix

\begin{figure*}
    \centering
    
    \subfloat[
    Macro F1  \\ (trained without DP = 0.60)]{\includegraphics[width=0.49\textwidth,trim={0.75cm 0 0 1.4cm},clip]{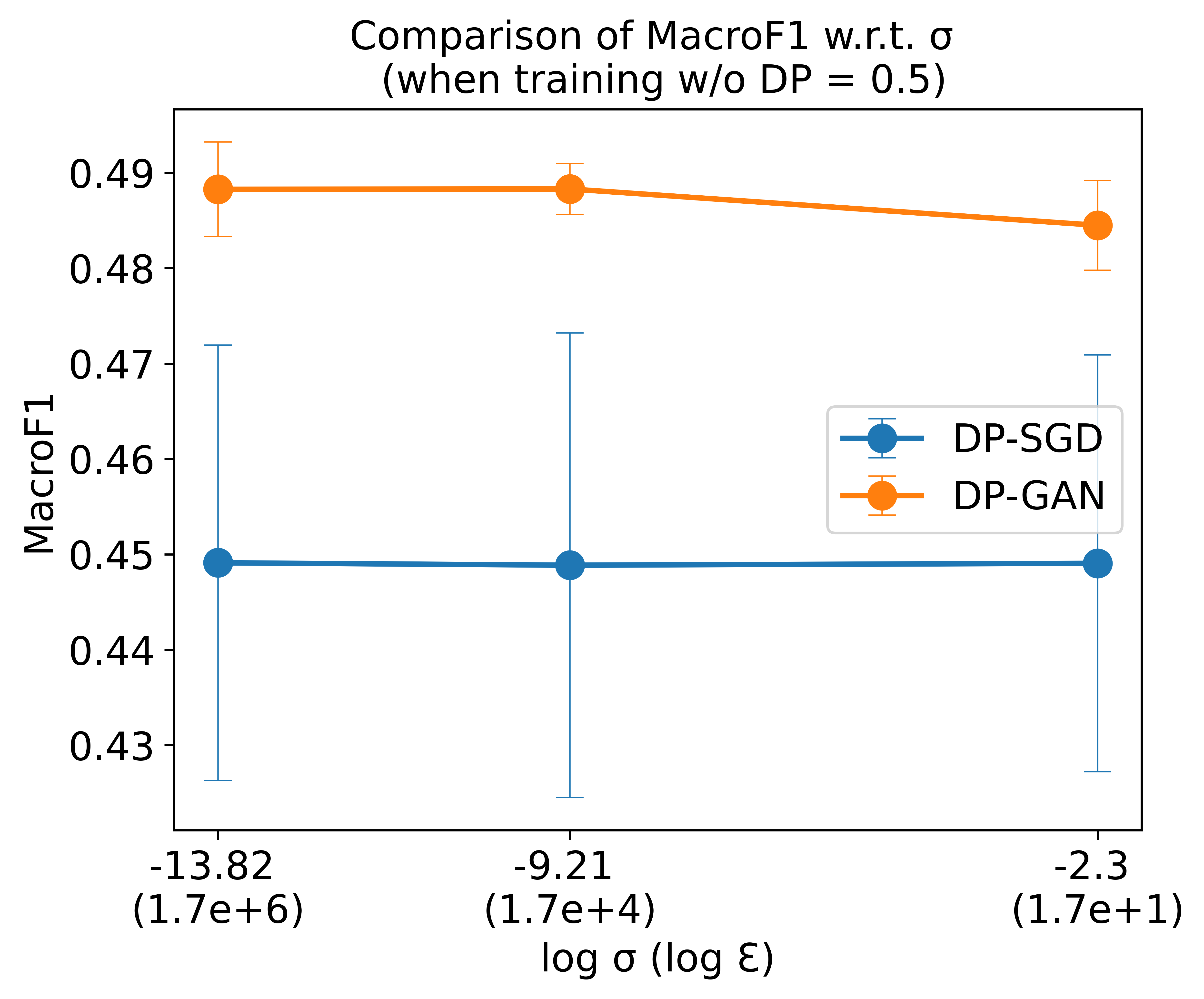}\label{fig6-a}}\hfill
    \subfloat[Euclidean Distance \\(trained without DP = 0.33)]{\includegraphics[width=0.49\textwidth,trim={0.75cm 0 0 1.4cm},clip]{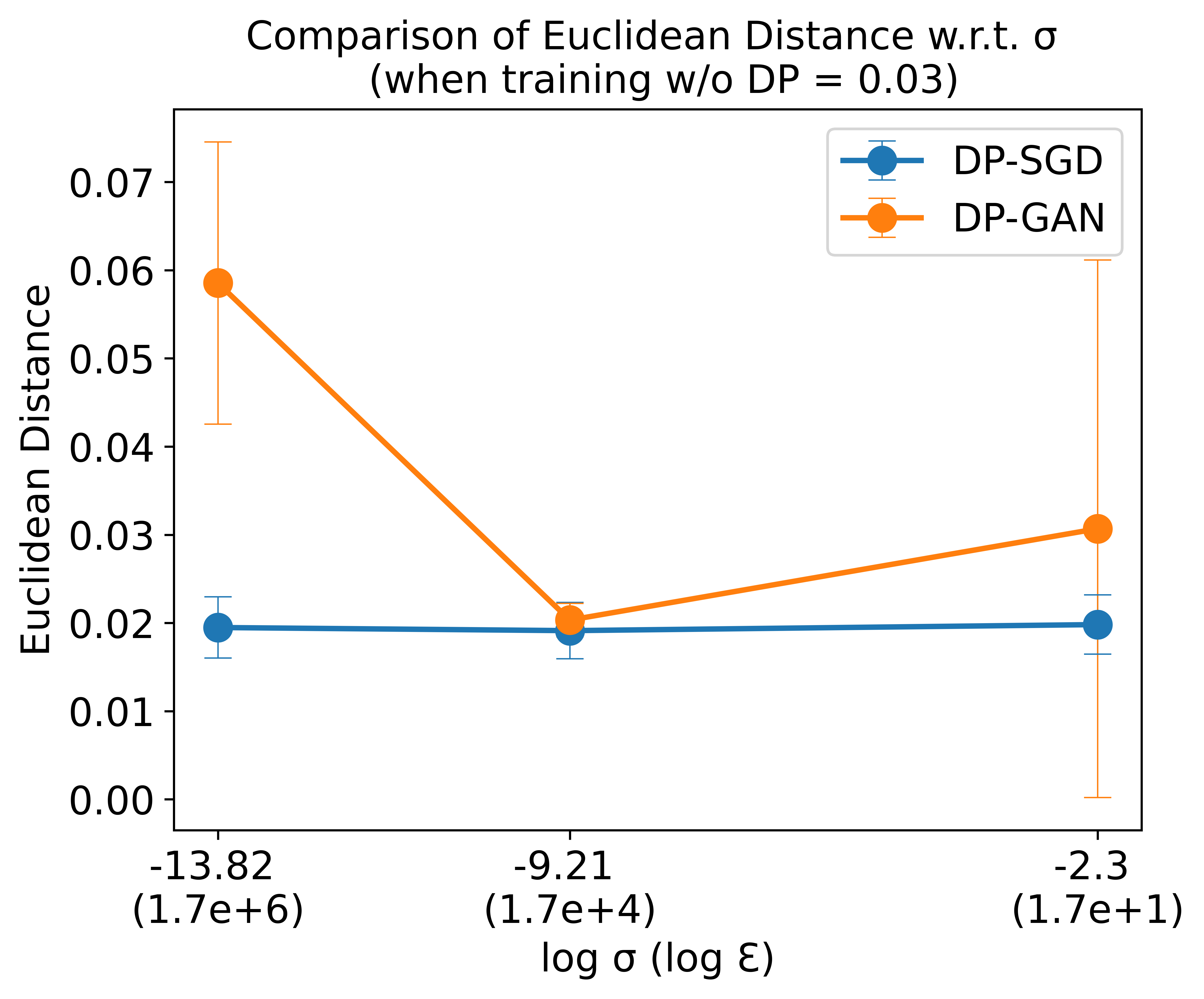}\label{fig6-b}} \\
    \subfloat[AUROC of the black-box attack \\ (trained without DP = 0.77)]{ \includegraphics[width=0.49\textwidth,trim={0.75cm 0 0 1.4cm},clip]{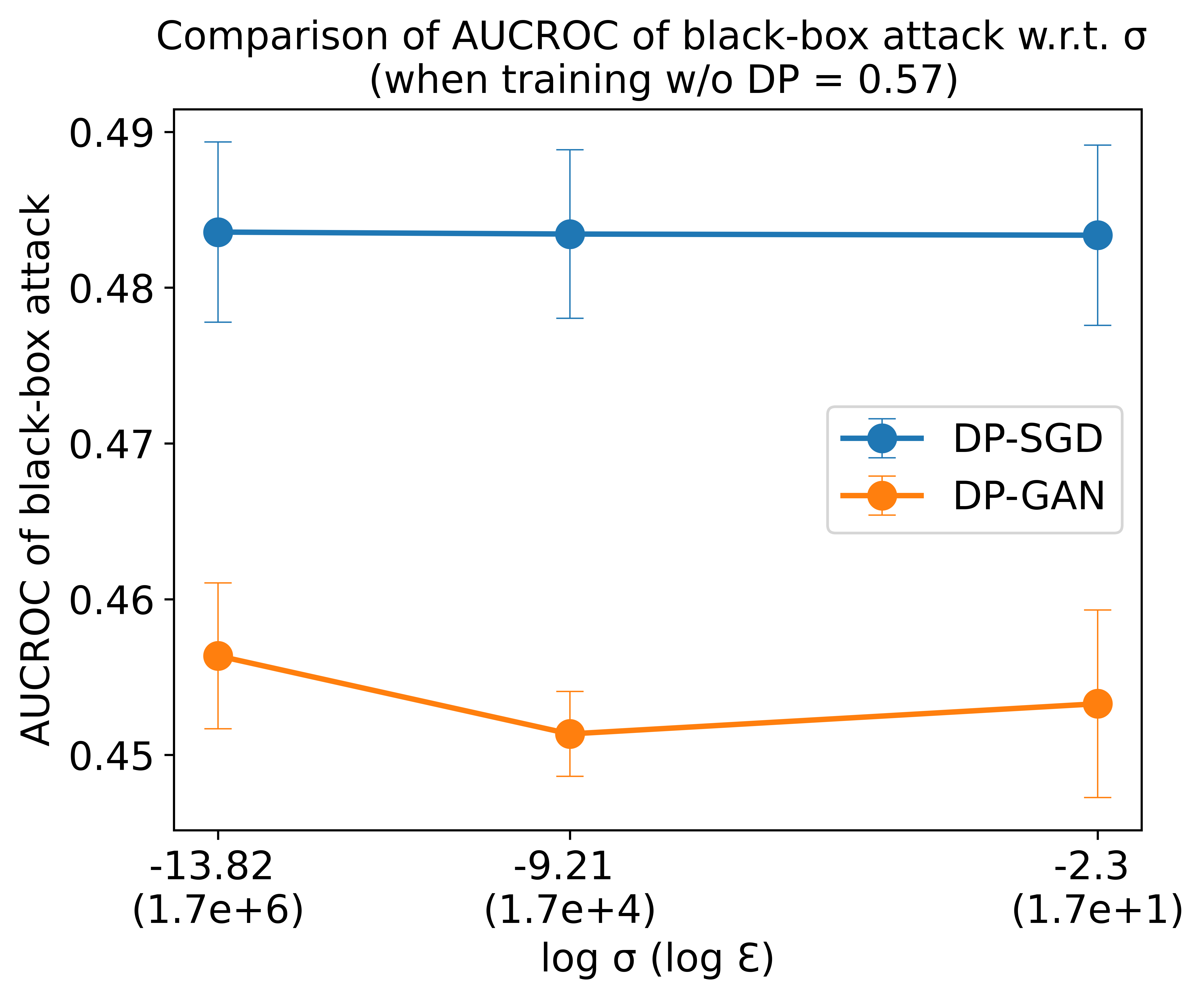}\label{fig6-c}}\hfill
    \subfloat[AUROC of the white-box attack \\ (trained without DP = 0.79)]{\includegraphics[width=0.49\textwidth,trim={0.75cm 0 0 1.4cm},clip]{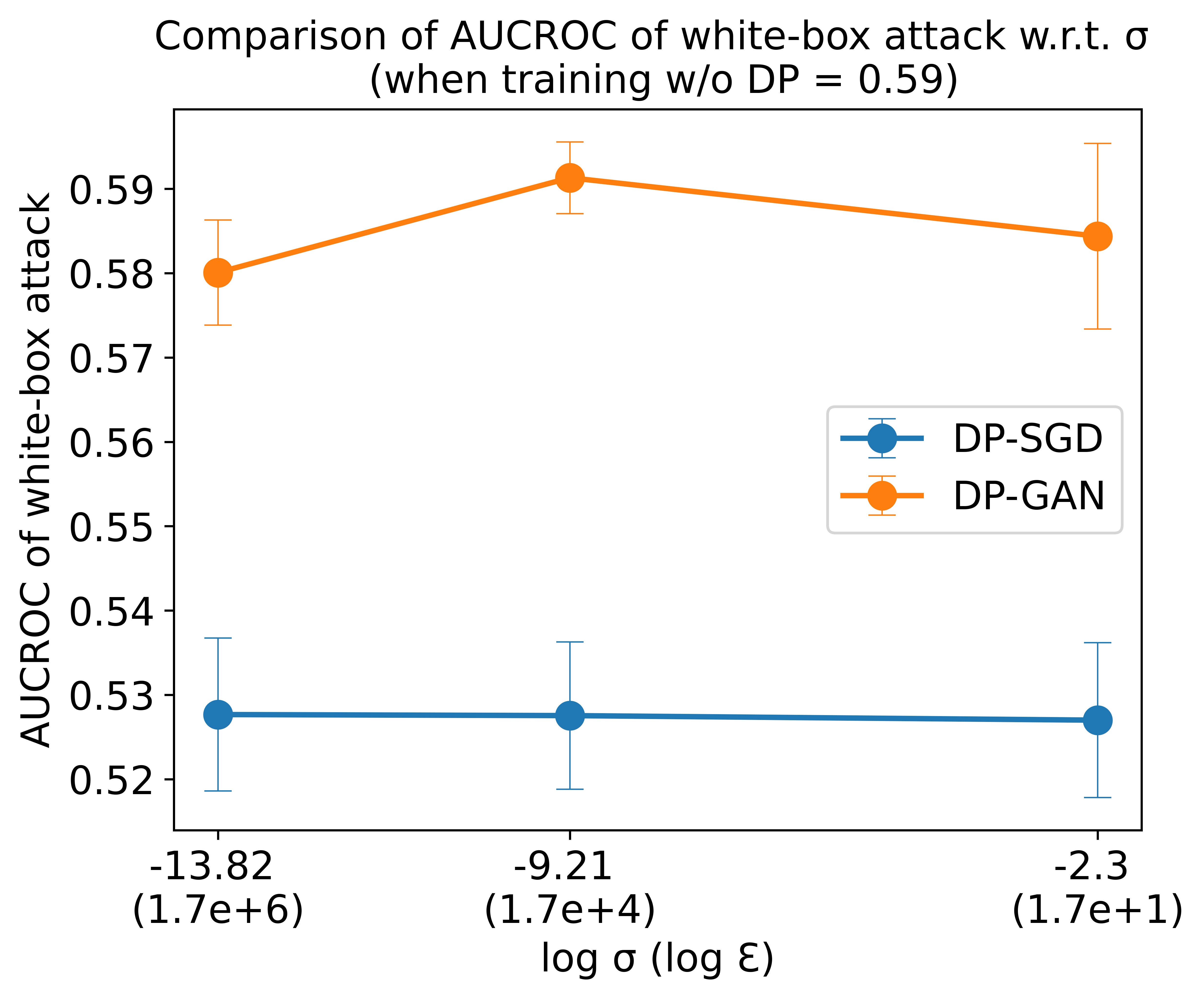}\label{fig6-d}}
            
    \caption{Comparisons between Macro F1, Euclidean distance and AUROC scores of the attacks on \texttt{CTGAN} for \texttt{Alphabank} trained with DP-SGD and DP-GAN w.r.t. {$\sigma=\{1e-05, 1e-04, 0.1\}$}, tested on 100 fake records. Values in the parentheses under $\sigma$ on X-axis mean $\epsilon$. $\sigma$ is log-transformed for better interpretation. Values in parentheses contain the results of the model trained without DP algorithms. For DP-GAN, we clip weights to be 0.005, a carefully selected value after the preceding experiment to decide an appropriate range of weight across the data sets. Note that in (b), the result has less Euclidean distance therefore, higher AUROC scores at $\log \sigma = -9.21$ than $\log \sigma = -2.3$, but still shows the same pattern as the other datasets in general, i.e., DP-GAN more preserves the generation quality of the models than DP-SGD, and it shows more robustness to the black-box attack while more vulnarability to the white-box attack.}
    \label{fig:fig6}
\end{figure*}

\begin{figure*}
    \centering
    
    \subfloat[
    $R^2$  \\ (trained without DP = 0.60)]{\includegraphics[width=0.49\textwidth,trim={0.75cm 0 0 0},clip]{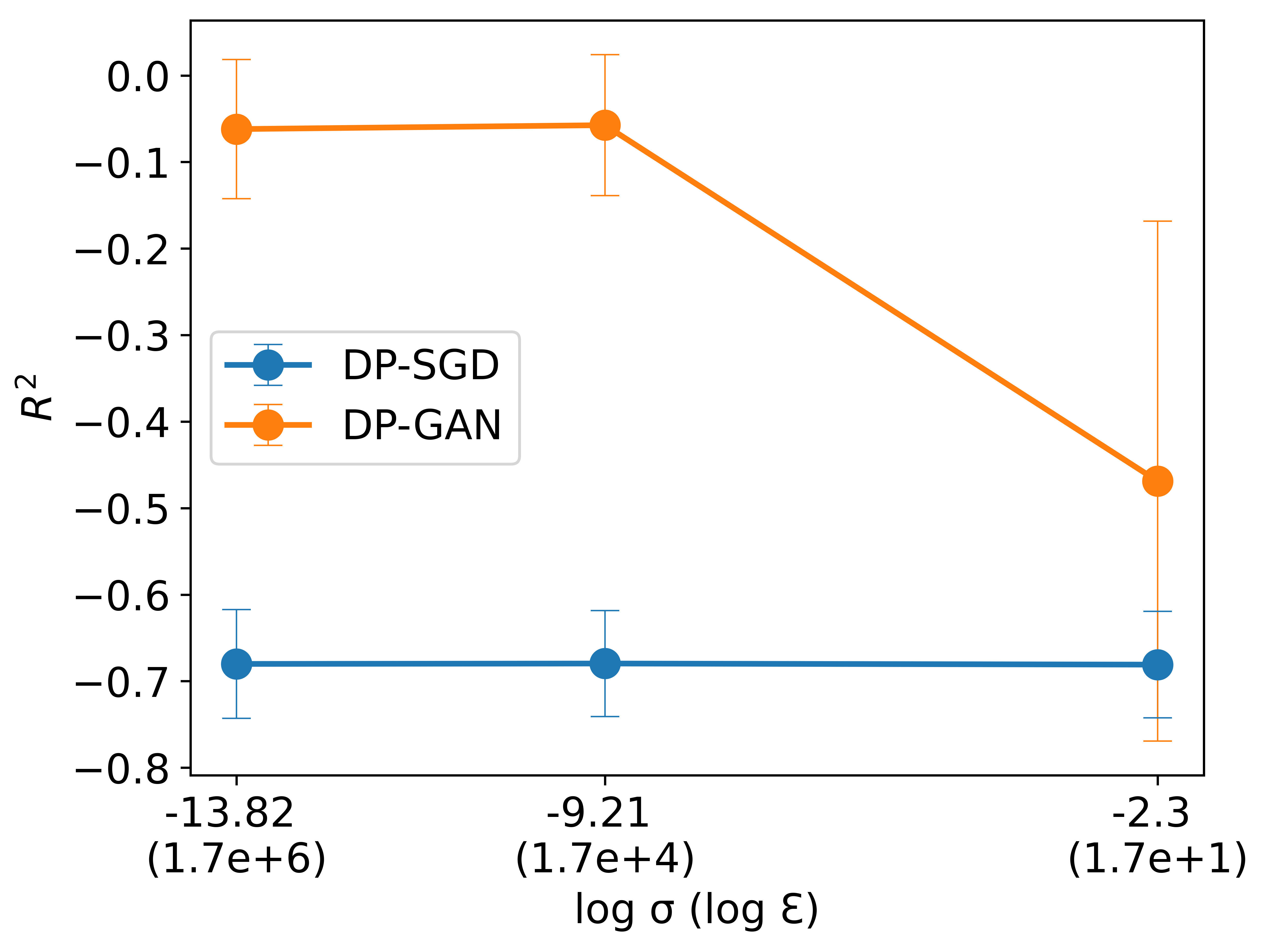}\label{fig7-a}}\hfill
    \subfloat[Euclidean Distance \\(trained without DP = 0.33)]{\includegraphics[width=0.49\textwidth,trim={0.75cm 0 0 0},clip]{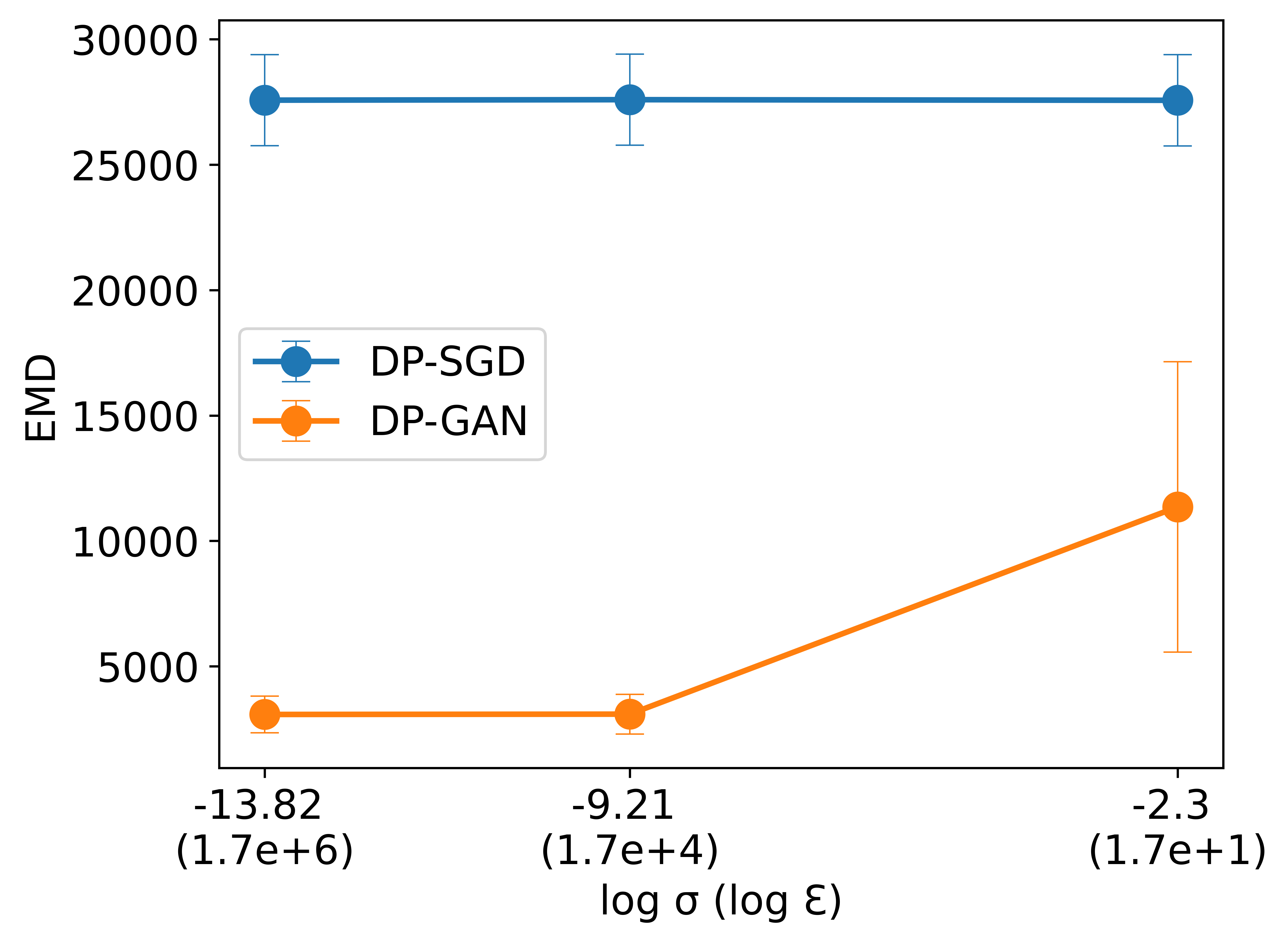}\label{fig7-b}} \\
    \subfloat[AUROC of the black-box attack \\ (trained without DP = 0.77)]{ \includegraphics[width=0.49\textwidth,trim={0.75cm 0 0 0},clip]{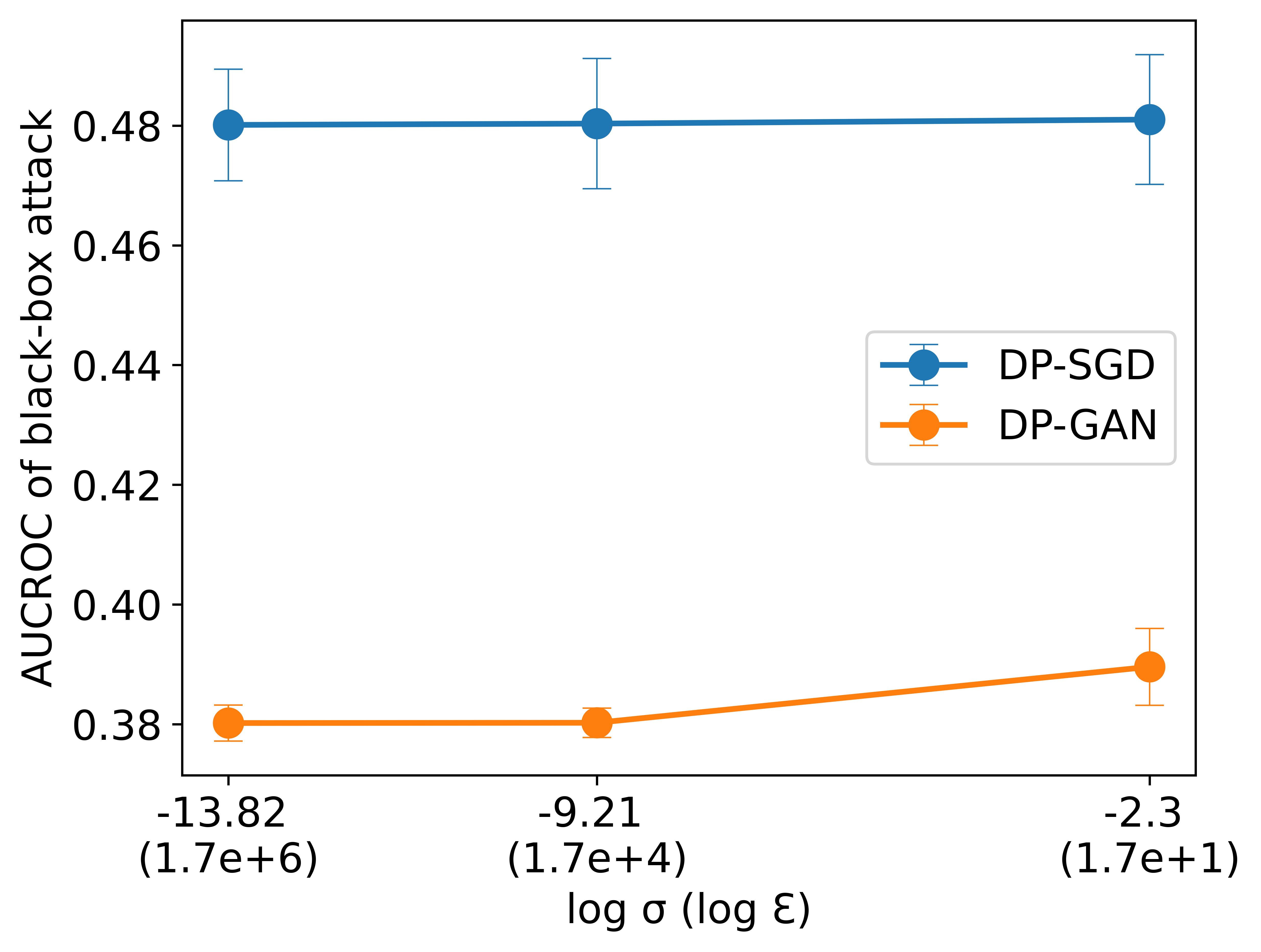}\label{fig7-c}}\hfill
    \subfloat[AUROC of the white-box attack \\ (trained without DP = 0.79)]{\includegraphics[width=0.49\textwidth,trim={0.75cm 0 0 0},clip]{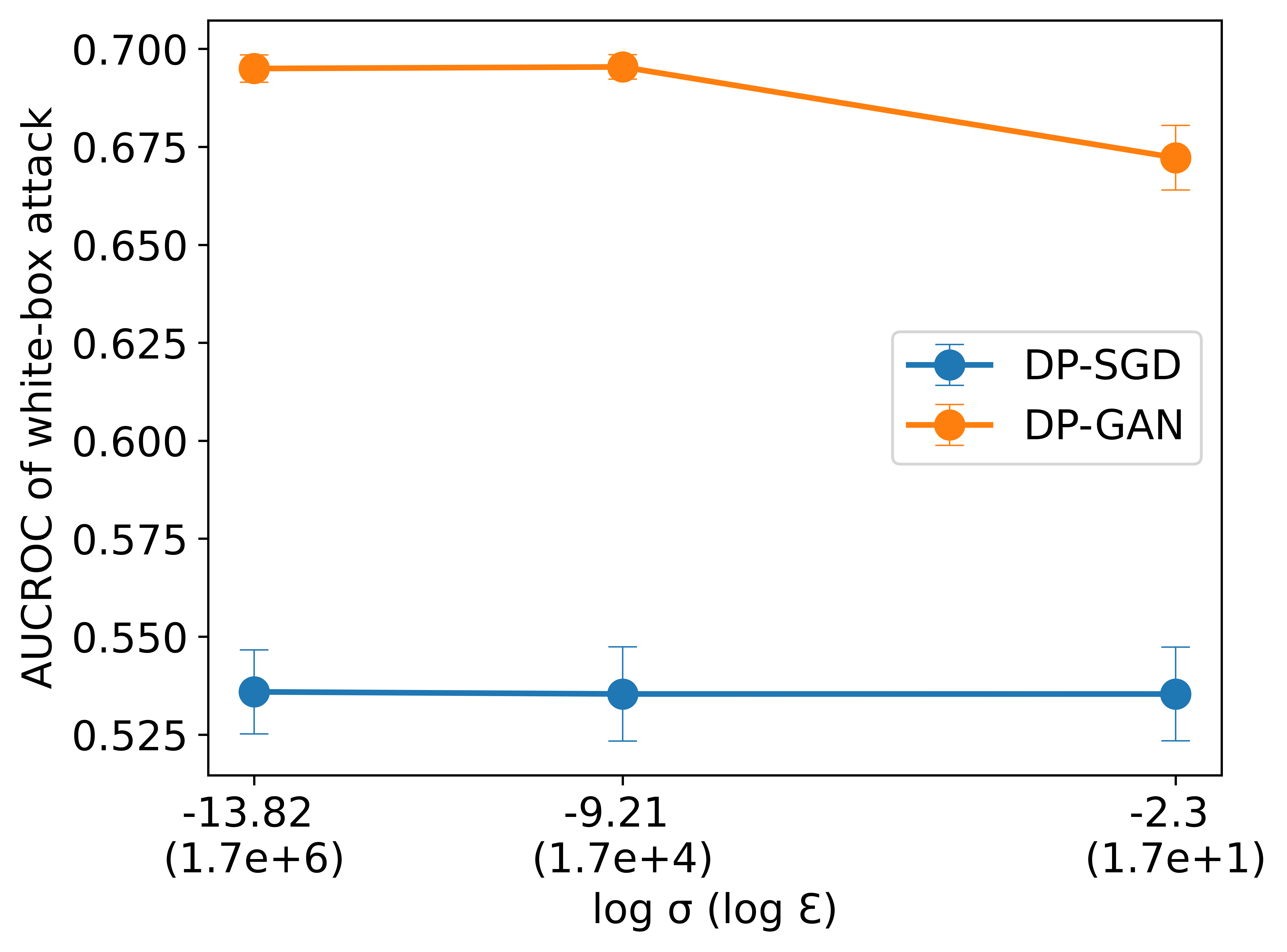}\label{fig7-d}}
            
    \caption{Comparisons between the generation quality and the attack scores on \texttt{CTGAN} for \texttt{King} trained with DP-SGD and DP-GAN w.r.t. {$\sigma=\{1e-06, 1e-04, 0.1\}$}, tested on 100 fake records. Values in the parentheses under $\sigma$ on X-axis mean $\epsilon$. For DP-GAN, we clip weights to be 0.01. We notice the tendency shown in \texttt{Adult} is observed in \texttt{King} as well; as the $\sigma$ increases, the Euclidean distance and AUROC of the black-box attack also increase, and AUROC of the white-box attack decreases. }
    \label{fig:fig7}
\end{figure*}

\begin{figure*}
    \centering
    
    \subfloat[
    Macro F1  \\ (trained without DP = 0.60)]{\includegraphics[width=0.49\textwidth,trim={0.75cm 0 0 0},clip]{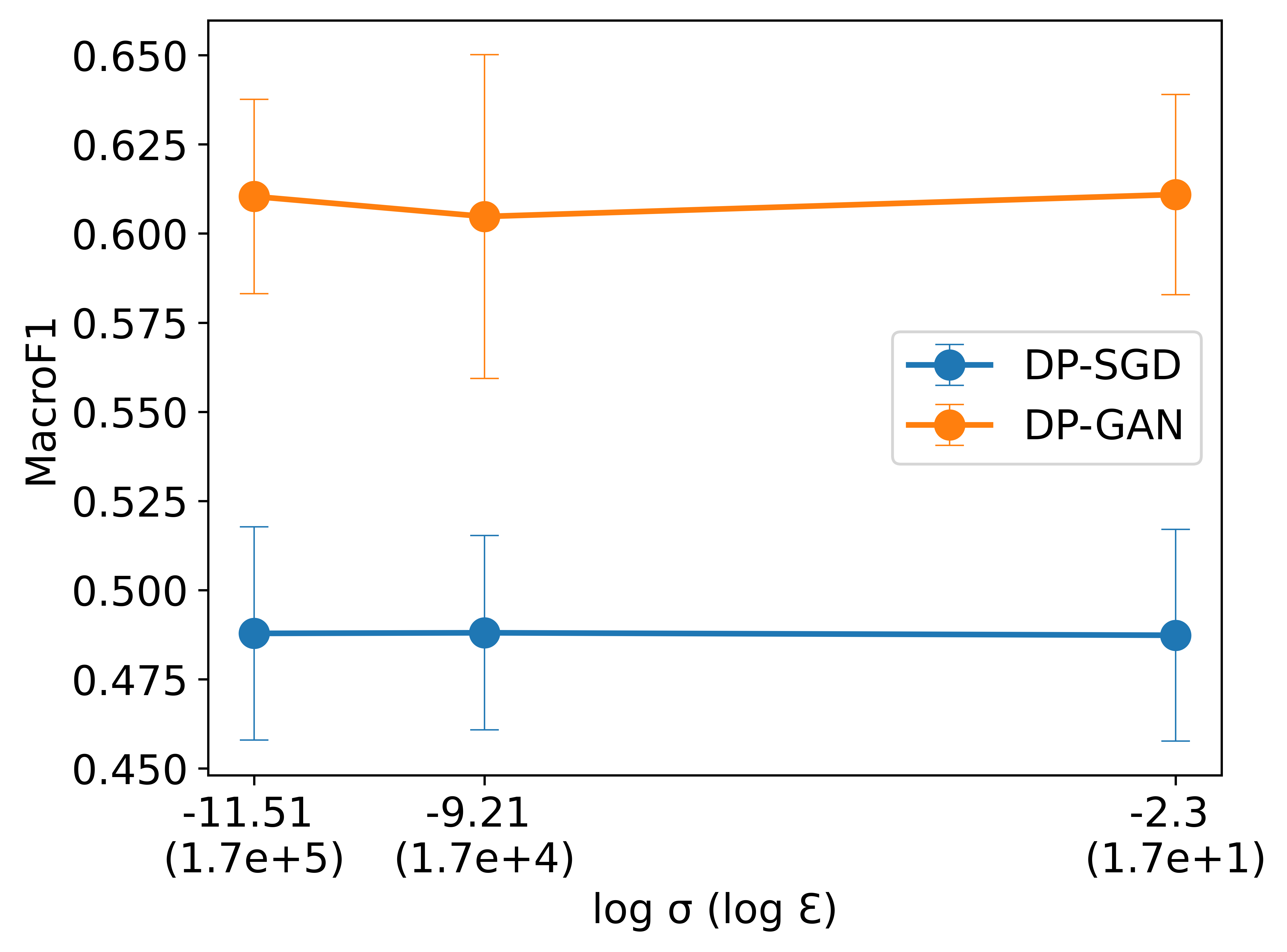}\label{fig8-a}}\hfill
    \subfloat[Euclidean Distance \\(trained without DP = 0.33)]{\includegraphics[width=0.49\textwidth,trim={0.75cm 0 0 0},clip]{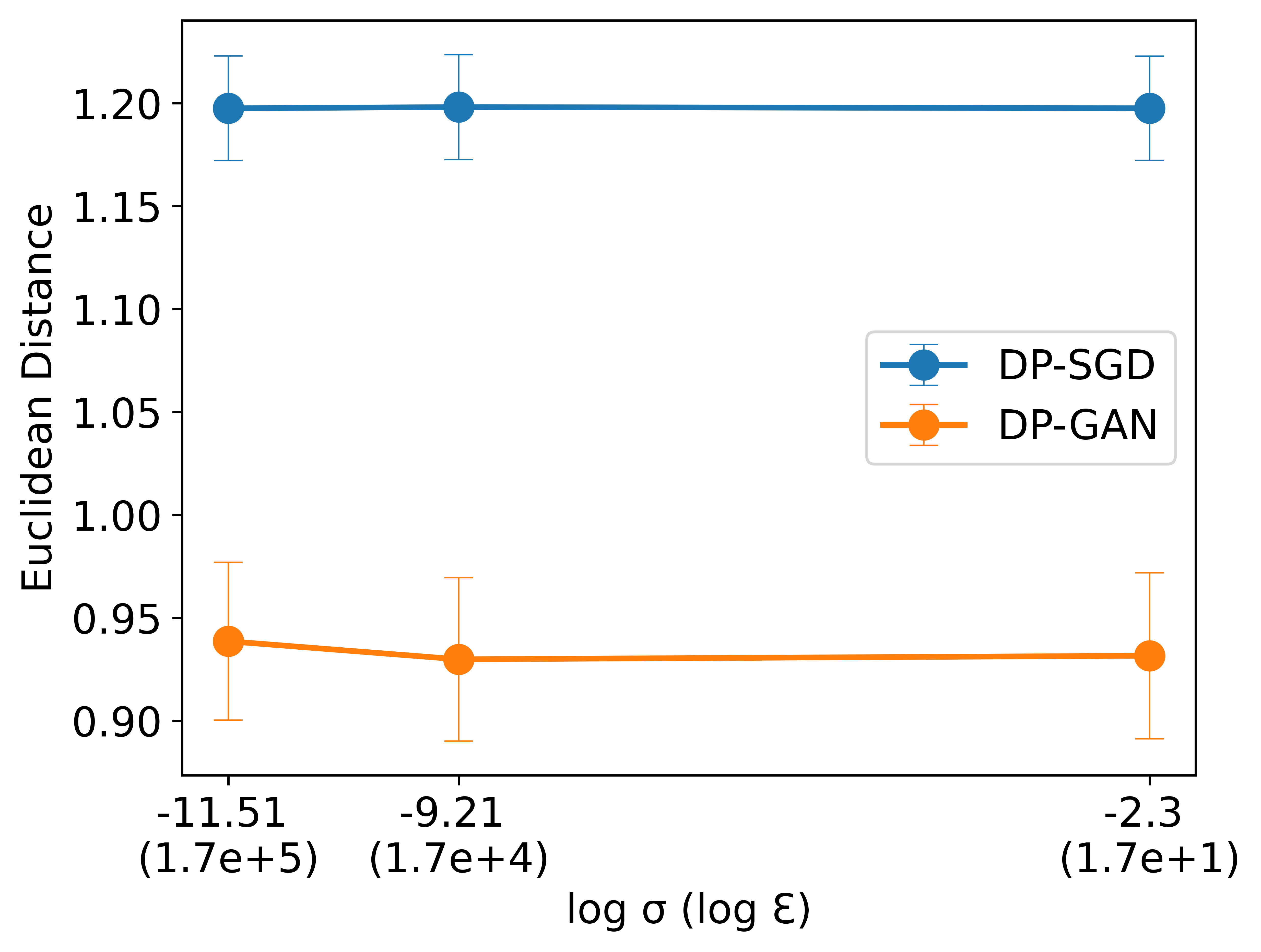}\label{fig8-b}} \\
    \subfloat[AUROC of the black-box attack \\ (trained without DP = 0.77)]{ \includegraphics[width=0.49\textwidth,trim={0.75cm 0 0 0},clip]{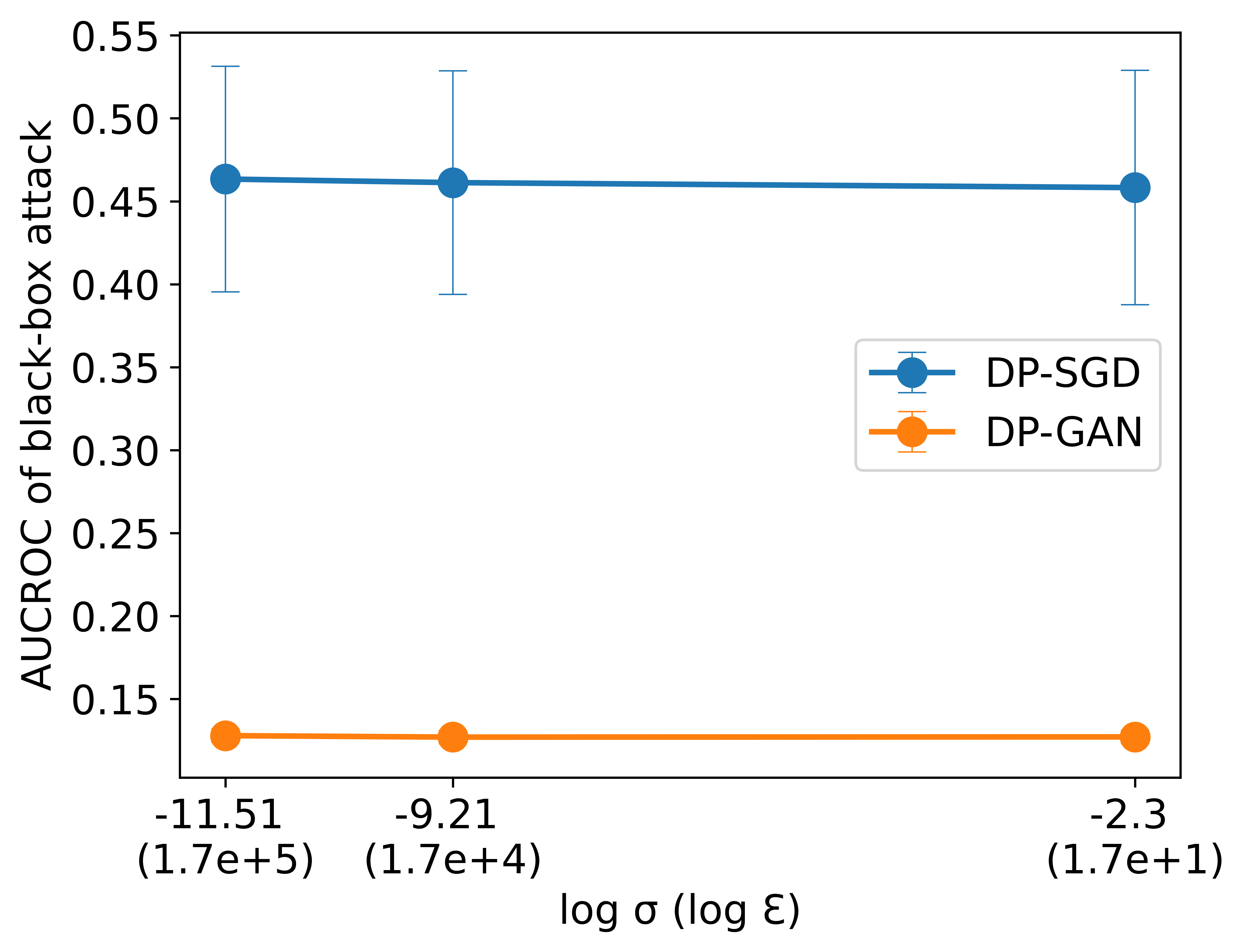}\label{fig8-c}}\hfill
    \subfloat[AUROC of the white-box attack \\ (trained without DP = 0.79)]{\includegraphics[width=0.49\textwidth,trim={0.75cm 0 0 0},clip]{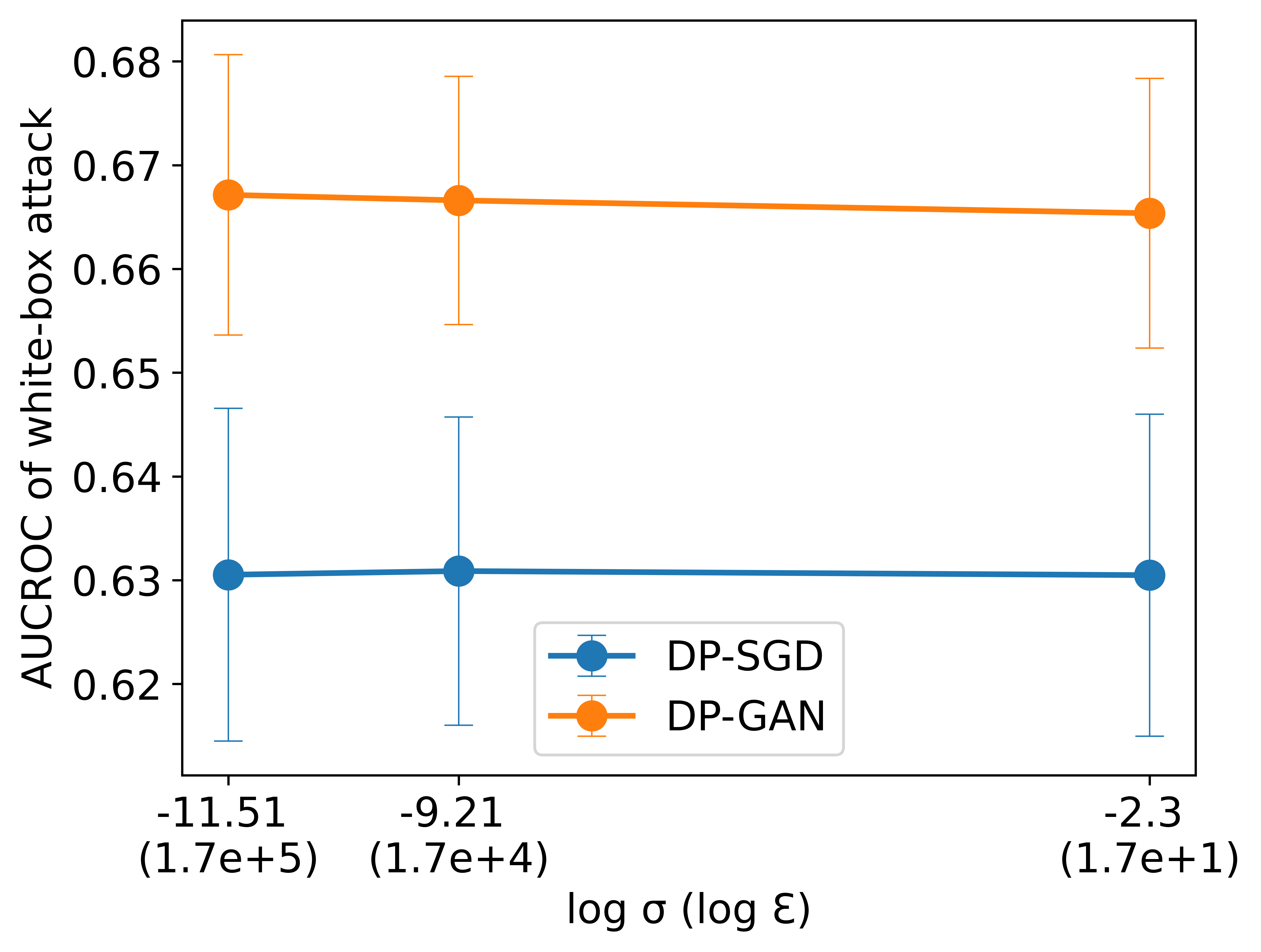}\label{fig8-d}}
            
    \caption{Comparisons between the generation quality and the attack scores on \texttt{CTGAN} for \texttt{Surgical} trained with DP-SGD and DP-GAN w.r.t. {$\sigma=\{1e-05, 1e-04, 0.1\}$}, tested with 100 fake records. Values in the parentheses under $\sigma$ on X-axis mean $\epsilon$. For DP-GAN, we clip weights to be 0.1. (d) The victim models are robust to the attacks for \texttt{Surgical} than the other data sets with under 0.67 of AUROC scores even with the DP-GAN algorithm when the AUROC of the model trained without DP is 0.79. Interestingly, the attack scores show a large standard deviation, which means \texttt{Surgical} is severely fluctuated on each attack. }
    \label{fig:fig8}
\end{figure*}

\end{document}